\documentclass[preprint,amsmath,amssymb,aps,showkeys,floatfix]{revtex4-2}
\usepackage[utf8]{inputenc}
\usepackage{hyperref}
\usepackage{braket}
\usepackage{bm}
\usepackage{dsfont}
\usepackage{graphicx}
\usepackage[export]{adjustbox}
\usepackage{simplewick}
\usepackage{subfigure}

\graphicspath{ {pictures/} }

\newcommand{\diagram}[1]{\hspace{2mm}\includegraphics[valign=c,scale=0.2]{#1}\hspace{2mm}}

\newcommand{\norm}[1]{\left\lVert#1\right\rVert}

\newcommand{\shortleftarrow}[1][3pt]{\mathrel{%
   \hbox{\usefont{U}{lasy}{m}{n}\symbol{40}}\hspace{-0.75mm}\hbox{\rule[\dimexpr\fontdimen22\textfont2-.2pt\relax]{#1}{.4pt}}%
   \mkern-4mu\phantom{\hbox{\usefont{U}{lasy}{m}{n}\symbol{40}}}}}

\begin{document}

\title{Interacting Scalar Field Theory on Causal Sets \footnote{Updated version of the draft submitted to the Handbook for Quantum Gravity (Springer).}}
\author{Ian Jubb}
\affiliation{Dublin Institute for Advanced Studies}
\date{\today}

\begin{abstract}
We introduce $\phi^4$ interacting real scalar Quantum Field Theory (QFT) on causal sets. We consider both the canonical framework of causal set free QFT, involving a Hilbert space and operators and so on, and the double path integral framework of causal set QFT outlaid by Sorkin. In both cases we describe how to extend the formalism to include a $\phi^4$ self-interaction, and, to make contact with the continuum, we contrast certain key expressions with their continuum counterparts. We develop a diagram-based algorithm, analogous to Feynman diagrams in the continuum, to compute the interacting 2-point function of our causal set QFT. Notably, causality is manifest in our diagrams in a manner not present in the usual Feynman diagrams of the continuum theory.
\end{abstract}

\maketitle
\newpage




\section{Introduction}

For the sake of consistency of causal set quantum gravity with the standard model - the latter being described through the framework of Quantum Field Theory (QFT) in a fixed background spacetime - it is imperative that we develop a theory of interacting quantum fields on causal sets, and one that agrees, at least in some regime, with observation~\footnote{That said, one could argue that disagreement at high energies could be interpreted as phenomenological evidence for causal set theory, though we will not explore this avenue here.}. In a full theory of causal set quantum gravity one imagines that spacetime itself would obey some quantum dynamics, and that there would be some back-reaction between the spacetime and any quantum fields in the spacetime. Without such a complete description at hand, and in the hopes that causal set quantum gravity can, in some regime, reproduce the standard model, we permit ourselves to first address quantum fields on a fixed background causal set~\footnote{When making comparisons to the continuum it is useful to also consider many causal sets realised via Poisson sprinklings into some given continuum spacetime.}.

Our current description of Quantum Field Theory (QFT) on a given fixed causal set is at a somewhat early stage of its development. Our lack of a description of tensors and/or spinors on causal sets limits us to only scalar fields, as opposed to fermionic or gauge fields. For this reason we focus here on interacting \textit{real scalar} quantum fields, and, for concreteness, consider only the specific case of a self-interacting $\phi^4$ theory. 

Textbook approaches to interacting QFT in the continuum usually begin with the canonical description of the free theory, involving Hilbert spaces and operators and so on, and then move over to the path integral framework when interactions are introduced~\cite{peskin1995introduction}. We will follow the same route here when introducing causal set interacting QFT. Before doing that, however, we recap some important aspects of the continuum theory in Section~\ref{sec:Continuum Framework}. Much of our discussion surrounding the path integral approach to causal set QFT will actually be more concerned with the \textit{double path integral}, or Schwinger-Keldysh formalism, as this will be more appropriate for our purposes. The textbook approach to path integral QFT is not often expressed via a double path integral, however, and hence we first devote some time to deriving some useful expressions in the continuum that will provide the basis/motivation for the analogous causal set double path integral expressions.

The canonical~\cite{stephen_johnston_thesis} and double path integral~\cite{Sorkin_scalar_field_histories} frameworks for \textit{free} scalar QFT on causal sets are recapped in Section~\ref{sec:Causal Set Free Scalar Field Theory}. Extensions to interacting fields have not yet been considered in detail in the literature. In~\cite{Sorkin_scalar_field_histories} Sorkin notes that the double path integral approach is particularly advantageous in that it offers a clear route to introducing a self-interaction, namely, by adding the appropriate $\phi^4$ term to the action~\cite{Sorkin_scalar_field_histories}. This is precisely what we do in Section~\ref{sec:Causal Set Interacting Scalar Field Theory}, and we find that, from the perspective of the canonical framework, one can interpret this modification to the action as a unitary transformation of the field operators. To finish, we focus on the 2-point function of the interacting theory in Section~\ref{sec:Interacting 2-point function}, and in Section~\ref{sec:The Analogue of Feynman diagrams} we derive some analogous Feynman rules for calculating it order by order in the interaction parameter. Our diagrams resemble those of the continuum theory, but with the added complication of two types of edges between vertices: directed and undirected. Our analogue Feynman rules are summarised for convenience in Section~\ref{sec:Summary of analogue Feynman diagrams and rules}.

While there is not any published work on the approach presented here to interacting QFT on causal sets, we highlight Albertini's thesis~\cite{emma_albertini_diss}, which is the inspiration for much of what follows. We note, however, that our approach to the causal set analogue of Feynman diagrams differs from that in~\cite{emma_albertini_diss}. It is important to also highlight the work in~\cite{Dable_Heath_2020}, where interacting QFT on causal sets is discussed through the lens of deformation quantisation. This method is somewhat more involved than the textbook approach considered here, and it would be of interest to determine if the two could be reconciled.

\section{Continuum Framework}\label{sec:Continuum Framework}

\subsection{Free theory}\label{sec:Free theory}

We first review some relevant concepts from real scalar QFT in a continuum spacetime, and derive some useful expressions in the double path integral framework.

Consider first the free theory. An important quantity for our purposes, and indeed for much of QFT, is the time-ordered correlation function,
\begin{equation}\label{eq:2 point time ordered correlation function}
    \bra{0} \phi^{(H)}(x_2) \phi^{(H)}(x_1) \ket{0}
\end{equation}
where $x_1$ and $x_2$ are two spacetime points, with time coordinates ordered as $x_1^0 < x_2^0$. In what follows, spatial coordinates of a point $x$ will be denoted as $\vec{x}$. $\ket{0}$ denotes the ground state of the theory, and $\phi^{(H)}(x) = U(0, x^0)^{\dagger}\phi^{(S)}(\vec{x})U(0, x^0)$ is the field operator~\footnote{Technically, $\phi^{(H)}(x)$ is an operator-valued distribution that one must integrate against a test function to yield a well-defined operator on the Hilbert space.} at $x$ in the Heisenberg picture. The Schr\"{o}dinger picture field, $\phi^{(S)}(\vec{x})$, is only a function of the spatial coordinates. Here $U(t,t')$ is the unitary operator that evolves the system from $t$ to $t'$.

In Peskin and Schroeder~\cite{peskin1995introduction}, they derive the path integral representation of~\eqref{eq:2 point time ordered correlation function}:
\begin{equation}\label{eq:peskin path integral}
    \bra{0} \phi^{(H)}(x_2) \phi^{(H)}(x_1) \ket{0} = \lim_{T\rightarrow \infty (1-i \epsilon)} \frac{\int \mathcal{D}\xi \, \xi(x_1)\xi(x_2)e^{i S[\xi]}}{\int \mathcal{D}\xi \, e^{i S[\xi]}} \; ,
\end{equation}
where $S[\xi] = \int_{-T}^T d^d x \, \mathcal{L}$ is the action from time $-T$ to $T$, and the integrals are over all classical \textit{spacetime} field configurations $\xi(x)\equiv \xi (x^0 , \vec{x})$ with the boundary conditions $\xi(\pm 
T,\vec{x}) = \zeta_{\pm}(\vec{x})$, where $\zeta_{\pm}(\vec{x})$ are some functions of the spatial coordinates only. The specific forms of the boundary functions $\zeta_{\pm}$ are actually irrelevant in the limit $T\rightarrow \infty (1-i \epsilon)$ (provided the support of the ground state wavefunctional includes these functions).

\vspace{2mm}
\noindent \textbf{Remark.} In the interacting theory, the act of taking $T\rightarrow\infty (1- i\epsilon )$ also ensures that one recovers the correlation function \textit{in the ground state of the interacting theory}. In the causal set case below we only consider finite causal sets, and hence we do not have the luxury of taking $T\rightarrow \infty$, let alone in a slightly imaginary direction. It is not clear, then, how one can recover the ground state of the interacting theory in the causal set case. We comment briefly on this at the end of Section~\ref{sec:Interacting 2-point function}, and for now content ourselves with the 2-point function in an arbitrary state $\ket{\Psi}$, i.e. $\bra{\Psi} \phi^{(H)}(x_2) \phi^{(H)}(x_1) \ket{\Psi}$.

\vspace{2mm}
In what follows the symbol $\xi$ will be used exclusively for functions of both space and time coordinates. For brevity, we also drop the $\vec{x}$ argument in $\xi(t,\vec{x})$, and simply write $\xi(t)$ when referring to the spatial function one obtains by evaluating $\xi$ at time $t$. Finally, the symbol $\zeta$ will be used exclusively for functions just of spatial coordinates. 

It will be convenient for our purposes to express $\bra{\Psi} \phi^{(H)}(x_2) \phi^{(H)}(x_1) \ket{\Psi}$ not as a single path integral as in~\eqref{eq:peskin path integral}, but as a \textit{double path integral}. To do this we first express the fields in the Schr\"{o}dinger picture. We have
\begin{equation}\label{eq:schro picture 2 point time ordered correlation function}
    \bra{\Psi} \phi^{(H)}(x_2) \phi^{(H)}(x_1) \ket{\Psi} = \bra{\Psi}U(0,T)^{\dagger} U(x_2^0,T)\phi^{(S)}(\vec{x}_2)U(x_1^0,x_2^0) \phi^{(S)}(\vec{x}_1) U(0,x_1^0)\ket{\Psi} \; ,
\end{equation}
for some arbitrary $T>x_{1,2}^0$. We next express the bra and kets in~\eqref{eq:schro picture 2 point time ordered correlation function} as
\begin{equation}\label{eq:gs wavefunctional}
    \bra{\Psi} = \int \mathcal{D}\zeta' \, \Psi(\zeta' )^* \bra{\zeta' } \;\; , \;\;\; \ket{\Psi} = \int \mathcal{D}\zeta \, \Psi(\zeta) \ket{\zeta } \; ,
\end{equation}
where the integrals are over all spatial configurations $\zeta\equiv \zeta(\vec{x})$, and $\Psi$ is the wavefunctional that returns a complex number for every spatial configuration $\zeta$.

In terms of this field configuration basis we similarly express the Schr\"{o}dinger field operators in~\eqref{eq:schro picture 2 point time ordered correlation function} as
\begin{equation}\label{eq:schro field operator in terms of zeta states}
    \phi^{(S)}(\vec{x}_2) = \int  \mathcal{D}\zeta_2  \, \zeta_2(\vec{x}_2)  \,\ket{\zeta_2}\bra{\zeta_2} \;\; , \;\;\; \phi^{(S)}(\vec{x}_1) = \int  \mathcal{D}\zeta_1  \, \zeta_1(\vec{x}_1)  \,\ket{\zeta_1}\bra{\zeta_1} \; .
\end{equation}
Finally, we insert the identity,
\begin{equation}\label{eq:identity in terms of zeta states}
    \mathds{1} = \int \mathcal{D} \zeta_T \,  \ket{\zeta_T}\bra{\zeta_T} \; ,
\end{equation}
in between $U(0,T)^{\dagger}$ and $U(x_2^0,T)$ in~\eqref{eq:schro picture 2 point time ordered correlation function}. We then have
\begin{align}\label{eq:integral over spatial configs}
    \bra{\Psi} \phi^{(H)}(x_2) \phi^{(H)}(x_1) \ket{\Psi} = & \int \mathcal{D}\zeta' \, \Psi(\zeta')^* \bra{\zeta'} U(0,T)^{\dagger} \int\mathcal{D}\zeta_T \, \ket{\zeta_T}\bra{\zeta_T} U(x_2^0,T) \nonumber
    \\
    & \times \int\mathcal{D}\zeta_2 \, \zeta_2(\vec{x}_2) \ket{\zeta_2}\bra{\zeta_2} U(x_1^0 , x_2^0) \int\mathcal{D}\zeta_1 \, \zeta_1(\vec{x}_1) \ket{\zeta_1}\bra{\zeta_1} \nonumber
    \\
    & \times U(0,x_1^0) \int\mathcal{D}\zeta \, \Psi(\zeta) \ket{\zeta} \nonumber
    \\
    = & \int \mathcal{D}\zeta' \mathcal{D}\zeta_T \mathcal{D}\zeta_2 \mathcal{D}\zeta_1 \mathcal{D}\zeta \, \Psi(\zeta')^* \Psi(\zeta) \bra{\zeta'}U(0,T)^{\dagger}\ket{\zeta_T}\nonumber
    \\
    & \times \bra{\zeta_T}U(x_2^0,T)\ket{\zeta_2} \bra{\zeta_2}U(x_1^0,x_2^0)\ket{\zeta_1}\bra{\zeta_1}U(x_1^0,0)\ket{\zeta} \nonumber
    \\
    & \times \zeta_1(\vec{x}_1)\, \zeta_2(\vec{x}_2) \; .
\end{align}
Next, we can rewrite each of the unitary transition amplitudes using the following identity~\cite{peskin1995introduction}
\begin{equation}\label{eq:unitary transition amplitude identity}
    \bra{\zeta'}U(t,t')\ket{\zeta} = \int_{\substack{\xi(t') = \zeta' \\ \xi(t) = \zeta}} \mathcal{D}\xi e^{iS[\xi]} \; ,
\end{equation}
where the integral is over all spacetime field configurations $\xi \equiv \xi(x)$ between the times $t$ and $t'$, with boundary values fixed on those time slices to the spatial functions $\zeta$ and $\zeta'$ respectively. Note that $S[\xi] = \int_t^{t'}d^dx \, \mathcal{L}$ denotes the action from $t$ to $t'$ in this case.

Using the identity~\eqref{eq:unitary transition amplitude identity}, the rhs of~\eqref{eq:integral over spatial configs} then becomes
\begin{equation}\label{eq:six integrals}
    \int \mathcal{D}\zeta' \mathcal{D}\zeta_T \mathcal{D}\zeta_2 \mathcal{D}\zeta_1 \mathcal{D}\zeta \int_{\substack{\overline{\xi}(T)=\zeta_T \\ \overline{\xi}(0)=\zeta_T}} \mathcal{D}\overline{\xi}  
    \int_{\substack{
    \xi(T) = \zeta_T \\
    \xi(x_2^0) = \zeta_2 \\
    \xi(x_1^0) = \zeta_1 \\
    \xi(0) = \zeta
    }}\mathcal{D}\xi \, e^{i ( S[\xi] - S[\overline{\xi}] )} \Psi(\zeta')^*\Psi(\zeta) \, \zeta_2 (\vec{x}_2) \zeta_1(\vec{x}_1) \; ,
\end{equation}
where the action is now from $0$ to $T$. As the spacetime field configuration $\xi$ is pinned to $\zeta_1$ at time $x_1^0$, and to $\zeta_2$ at time $x_2^0$, we can replace $\zeta_2 (\vec{x}_2) \zeta_1(\vec{x}_1)$ by $\xi(x_2)\xi(x_1)$ in the integrand. The six integrals in~\eqref{eq:six integrals} amount to a double path integral over all pairs of spacetime field configurations $\overline{\xi}$ and $\xi$, but with the restriction that the two configurations agree at time $T$. This restriction can be encoded via the delta function $\delta ( \xi(T) - \overline{\xi}(T) )$. We now have our final expression:
\begin{align}
    \bra{\Psi} \phi^{(H)}(x_2) \phi^{(H)}(x_1) \ket{\Psi} = \int \mathcal{D}\overline{\xi}\mathcal{D}\xi \, e^{i ( S[\xi] - S[\overline{\xi}] )} \delta ( \xi(T) - \overline{\xi}(T) ) \Psi(\overline{\xi}(0))^*\Psi(\xi(0)) \, \xi (x_2)\xi(x_1) \; .
\end{align}
We can rewrite this expression as
\begin{equation}\label{eq:2 point function as integral over free decoherence functional}
    \bra{\Psi} \phi^{(H)}(x_2) \phi^{(H)}(x_1)\ket{\Psi} = \int \mathcal{D}\overline{\xi}\mathcal{D}\xi \, D(\xi,\overline{\xi}) \, \xi (x_2)\xi(x_1) \; ,
\end{equation}
where we have introduced the \textit{decoherence functional}
\begin{equation}\label{eq:continuum decoherence functional form}
    D(\xi,\overline{\xi}) = e^{i ( S[\xi] - S[\overline{\xi}] )} \delta ( \xi(T) - \overline{\xi}(T) ) \Psi(\overline{\xi}(0))^*\Psi(\xi(0)) \; .
\end{equation}

\subsection{Interacting theory}

To move to the interacting theory, one modifies the double path integral expression by adding an interaction term to the action: $S[\xi]\mapsto S[\xi;\lambda] = S[\xi]+S_{\text{int}}[\xi;\lambda]$, where $S[\xi]$ is the action of the free theory, and $\lambda$ is the coupling or interaction, parameter. The interacting decoherence functional is then
\begin{align}\label{eq:continuum interacting decoherence functional form}
    D(\xi,\overline{\xi};\lambda) & = e^{i ( S[\xi;\lambda] - S[\overline{\xi};\lambda] )} \delta ( \xi(T) - \overline{\xi}(T) ) \Psi(\overline{\xi}(0))^*\Psi(\xi(0)) 
    \nonumber
    \\
    & = D(\xi , \overline{\xi})e^{i(S_{\text{int}}[\xi;\lambda]-S_{\text{int}}[\overline{\xi};\lambda])} \; .
\end{align}
Following the logic of Section~\ref{sec:Free theory} in reverse, if we compute the integral in~\eqref{eq:2 point function as integral over free decoherence functional} with $D(\xi,\overline{\xi})$ replaced by $D(\xi,\overline{\xi};\lambda)$, i.e. $\int \mathcal{D}\overline{\xi}\mathcal{D}\xi \, D(\xi,\overline{\xi}; \lambda) \, \xi (x_2)\xi(x_1)$, we find $\bra{\Psi}\phi^{(H)}(x_2)\phi^{(H)}(x_1)\ket{\Psi}$ as we did in the free case. It is important to note, however, that even though the free and interacting double path integrals amount to the same expression in terms of the Heisenberg picture fields, i.e. $\bra{\Psi}\phi^{(H)}(x_2)\phi^{(H)}(x_1)\ket{\Psi}$, the 2-point functions \textit{are} different due to differences in the Heisenberg picture fields themselves. In the free theory, one finds the Heisenberg fields by evolving the Schr\"{o}dinger picture fields with the free Hamiltonian, $H_0$, as $\phi^{(H)}(x) = e^{i H_0 x^0}\phi^{(S)}(\vec{x})e^{-iH_0 x^0}$, whereas in the interacting theory, they are evolved with the full Hamiltonian, $H(t) = H_0 + H_{int}(t)$, as $\phi^{(H)}(x) = T\lbrace e^{i\int_0^{x^0}dt H(t)} \rbrace\phi^{(S)}(\vec{x})T\lbrace e^{-i\int_0^{x^0}dt H(t)} \rbrace$. In making comparisons between the causal set and continuum field theories below, it will be more convenient to focus on the \textit{interaction picture} fields in the interacting theory - those that are evolved with the free Hamiltonian as $\phi^{(I)}(x) = e^{i H_0 x^0}\phi^{(S)}(\vec{x})e^{-iH_0 x^0}$. It should be clear from their definition that they are the same as the Heisenberg fields in the free theory. Henceforth, any use of $\phi(x)$ without a superscript should be understood as an interaction picture field, unless otherwise stated.

In terms of the interaction picture fields one can verify that
\begin{equation}\label{eq:2 point function continuum interacting theory}
    \bra{\Psi}\phi^{(H)}(x_2)\phi^{(H)}(x_1)\ket{\Psi} = \bra{\Psi}V(0,T)^{\dagger}V(x_2^0,T)\phi^{(I)}(x_2)V(x_1^0,x_2^0)\phi^{(I)}(x_1)V(0,x_1^0)\ket{\Psi} \; ,
\end{equation}
where 
\begin{equation}\label{eq:interaction unitary definition}
    V(t,t') = T\lbrace \exp ( -i \int_t^{t'} dt'' H_{int}^{(I)}(t'') ) \rbrace \; ,
\end{equation}
and $H_{int}^{(I)}(t) = e^{i H_0 t}H_{int}(t)e^{-iH_0 t}$ is the interacting part of the full Hamiltonian expressed in the interaction picture. For a self-interacting theory, say $\phi^4$, this would take the form of an integral over some spatial surface $\Sigma$ of constant time $t$:
\begin{equation}\label{eq:interaction hamiltonian definition}
    H_{int}^{(I)}(t) = -\int_{\Sigma}d^{d-1}\vec{x}\, \sqrt{h(x)} \, \frac{\lambda}{4!}\, \phi(x)^4 \; ,
\end{equation}
where $x=(t,\vec{x})$, and where $\phi(x)\equiv \phi^{(I)}(x)$ is in the interaction picture. Here $h(x)$ is the determinant of metric induced on $\Sigma$, evaluated at $x$, and $\lambda$ is the coupling parameter.

Before moving on, we note that to recover higher $n$-point functions, e.g. $\bra{\Psi}\phi(x_n)...\phi(x_1)\ket{\Psi}$, from the double path integral framework, one simply inserts $\xi(x_i)$ in the double path integrand for each $i=1,...,n$. What may seem somewhat redundant at this point is the inclusion of the path integral over field configurations $\overline{\xi}$, as in any computation of an $n$-point function it seems we simply integrate out this entire field configuration. In a moment we will see that, if we ask for more than $n$-point correlation functions - specifically for probabilities of measurement outcomes - then this second path integral arises naturally.

\subsection{Expressing the decoherence functional in the canonical framework}\label{sec:Expressing the decoherence functional in the canonical framework}

To motivate our causal set expressions in Sections~\ref{sec:Histories Framework} and~\ref{sec:Introducing an interaction} it will be helpful to re-express the decoherence functionals in~\eqref{eq:continuum decoherence functional form} and~\eqref{eq:continuum interacting decoherence functional form} in terms of the quantities of the canonical framework. This will be a somewhat formal discussion, as certain steps in this process are not well defined. In the causal set case, however, such steps are all well defined.

Let us say we want to calculate the probability (or more accurately the probability density) that a measurement of $\phi(x_1)$ at $x_1$ (so at time $x_1^0$ and position $\vec{x}_1$) gives a value of $\eta_1\mathbb{R}$, and that a measurement of $\phi(x_2)$ gives a value of $\eta_2\in\mathbb{R}$. To determine this probability we first evolve our initial state $\ket{\Psi}$ to time $x_1^0$: $\ket{\Psi}\mapsto U(0,x_1^0)\ket{\Psi}$. We then project with the (formal) Schr\"{o}dinger picture projector associated with the observation of $\eta_1$: $U(0,x_1^0)\ket{\Psi}\mapsto P_1^{(S)}U(0,x_1^0)\ket{\Psi}$, where $P_1^{(S)} := \int_{\zeta_1 = \eta_1} \mathcal{D}\zeta_1 \ket{\zeta_1 }\bra{\zeta_1}$, and where the integral is over all spatial field configurations $\zeta_1$ such that $\zeta_1(\vec{x}_1)=\eta_1$. We then evolve to $x_2^0$ with $U(x_1^0,x_2^0)$, and finally project with $P_2^{(S)}$, the associated projector for our observation of $\eta_2$. Our desired probability is then the modulus squared of this final state:
\begin{equation}
    \norm{ P^{(S)}_2 U(x_1^0,x_2^0) P^{(S)}_1 U(0,x_1^0) \ket{\Psi}}^2 = \norm{ P^{(H)}_2 P^{(H)}_1 \ket{\Psi}}^2 = \norm{ P^{(I)}_2 V(x_1^0,x_2^0) P^{(I)}_1 V(0,x_1^0) \ket{\Psi}}^2 \; ,
\end{equation}
where we have also included the Heisenberg and interaction picture expressions for convenience. Focusing on the interaction picture expression (as this will be directly comparable when we consider causal sets below) we can then derive the corresponding double path integral expression as follows:
\begin{align}\label{eq:prob of two measurements in interval path integral expression}
    & \norm{ P^{(I)}_2 V(x_1^0,x_2^0) P^{(I)}_1 V(0,x_1^0) \ket{\Psi}}^2  \nonumber
    \\
    & = \left( \bra{\Psi}V(0,x_1^0)^{\dagger}P^{(I)}_1 V(x_1^0,x_2^0)^{\dagger}P^{(I)}_2 \right) \left( P^{(I)}_2 V(x_1^0,x_2^0) P^{(I)}_1 V(0,x_1^0) \ket{\Psi} \right)
    \nonumber
    \\
    & =  \bra{\Psi}V(0,x_1^0)^{\dagger}P^{(I)}_1 V(x_1^0,x_2^0)^{\dagger}P^{(I)}_2 V(x_2^0 , T)^{\dagger} V(x_2^0 , T) P^{(I)}_2 V(x_1^0,x_2^0) P^{(I)}_1 V(0,x_1^0) \ket{\Psi} 
    \nonumber
    \\
    & = \int_{\substack{\overline{\xi}(x_1)= \eta_1 \\ \overline{\xi}(x_2)= \eta_2}} \mathcal{D}\overline{\xi}\int_{\substack{\xi(x_1)= \eta_1 \\ \xi(x_2)= \eta_2}}\mathcal{D}\xi \, D(\xi,\overline{\xi}\; \lambda) \; ,
\end{align}
where in the last line the two path integrals are over all field configurations $\xi$ and $\overline{\xi}$ which evaluate to $\eta_1$ at $x_1$ and $\eta_2$ at $x_2$ respectively. Here the two path integrals have arisen, in some sense, from the two brackets in the second line of~\eqref{eq:prob of two measurements in interval path integral expression}.

We can encode the restrictions on the path integrals via the delta functions $\delta(\xi(x_1)-\eta_1)\delta(\xi(x_2)-\eta_2)$, and similarly for $\overline{\xi}$. Following the logic of Section~\ref{sec:Free theory}, this allows us to reverse engineer the corresponding expression in the canonical framework:
\begin{align}\label{eq:two delta functions double path integral}
    \int_{\substack{\overline{\xi}(x_1)= \eta_1 \\ \overline{\xi}(x_2)= \eta_2}} & \mathcal{D} \overline{\xi}\int_{\substack{\xi(x_1)=\eta_1 \\ \xi(x_2)=\eta_2}}\mathcal{D}\xi \, D(\xi,\overline{\xi};\lambda) \nonumber
    \\
    & = \int \mathcal{D}\overline{\xi}\mathcal{D}\xi \, D(\xi,\overline{\xi};\lambda) \, \delta(\overline{\xi}(x_1)-\eta_1)\delta(\overline{\xi}(x_2)-\eta_2)\delta(\xi(x_2)-\eta_2) \delta(\xi(x_1)-\eta_1)
    \nonumber
    \\
    & = \bra{\Psi}V(0,x_1^0)^{\dagger}\delta(\phi(x_1)-\eta_1) V(x_1^0,x_2^0)^{\dagger}\delta(\phi(x_2)-\eta_2) V(x_2^0 , T)^{\dagger} 
    \nonumber
    \\
    & \hspace{5mm} \times V(x_2^0 , T) \delta(\phi(x_2)-\eta_2) V(x_1^0,x_2^0) \delta(\phi(x_1)-\eta_1) V(0,x_1^0) \ket{\Psi} \; ,
\end{align}
where we recall that by $\phi(x)$ we mean the interaction picture fields evolved with the free Hamiltonian. It is important to note here the resulting order of delta functions in the last line of~\eqref{eq:two delta functions double path integral}, as these delta functions cannot necessarily be reordered if the fields $\phi(x_1)$ and $\phi(x_2)$ do not commute. From left to right, we first have the reverse-time ordered product of $\delta(\phi(x_1) - \eta_1)$ and $\delta(\phi(x_2) - \eta_2)$, with the appropriate interaction picture unitaries, $V(\cdot,\cdot)^{\dagger}$, inserted in between. After this we then have the time ordered product of $\delta(\phi(x_1) - \eta_1)$ and $\delta(\phi(x_2) - \eta_2)$, again with appropriate insertions of $V(\cdot,\cdot)$.

This precise ordering of these delta functions, and the precise insertions of $V(\cdot , \cdot)^{\dagger}$ and $V(\cdot,\cdot)$ in between them, generalises to any finite number of spacetime points $x_1, ... , x_N$. We can further consider the case where the field configurations $\xi$ and $\overline{\xi}$ are pinned to different values at $x_i$, say $\xi(x_i)=\eta_i$ and $\overline{\xi}_i=\overline{\eta}_i$. 

In fact, in will be useful to write down the explicit expression for a lattice of spacetime points, as this will help motivate some of our causal set expressions in Sections~\ref{sec:Histories Framework} and~\ref{sec:Introducing an interaction}. Consider, then, a regular square lattice of points, $\Lambda$, embedded in a $d$-dimensional spacetime $M$, truncated in time from $t=0$ to $t=T$, and whose constant time surfaces, $\Sigma$, are of compact topology. Consider a regular lattice in $M$ with finite lattice spacing, and only a finite number of points, i.e. $|\Lambda| < \infty$. This ensures we do not have to deal with any troublesome infinities in what follows. For convenience, let the set of time coordinates of the points in our lattice be $\mathcal{T} = \lbrace \Delta t,2\Delta t, ..., T-\Delta t\rbrace$, where $\Delta t$ is the lattice spacing in time. Let us also denote the set of all spatial coordinates of our lattice points as $\mathcal{X}$. We assume our lattice can then be expressed as the Cartesian product $\Lambda = \mathcal{T}\times \mathcal{X}$.

Now, consider two spacetime field configurations $\eta(x)$ and $\overline{\eta}(x)$ which we will pin $\xi$ and $\overline{\xi}$ to respectively, for each point in our lattice. We then have
\begin{align}\label{eq:lattice delta functions double path integral}
    \int_{\substack{\overline{\xi}(x)=\overline{\eta}(x), \\ \forall x\in\Lambda}}\mathcal{D} \overline{\xi} & \int_{\substack{\xi(x)=\eta(x), \\ \forall x\in\Lambda}} \mathcal{D}\xi \, D(\xi,\overline{\xi};\lambda)
    \nonumber
    \\
    & = \int \mathcal{D}\overline{\xi}\mathcal{D}\xi \, D(\xi,\overline{\xi};\lambda) \, 
    \prod_{x\in\Lambda} \delta(\overline{\xi}(x)-\overline{\eta}(x))\delta(\xi(x)-\eta(x))
    \nonumber
    \\
    & = \int \mathcal{D}\overline{\xi}\mathcal{D}\xi \, D(\xi,\overline{\xi};\lambda) \, 
    \prod_{t\in\mathcal{T}}\prod_{\vec{x}\in\mathcal{X}} \delta(\overline{\xi}(t,\vec{x})-\overline{\eta}(t,\vec{x}))\delta(\xi(t,\vec{x})-\eta(t,\vec{x}))
    \nonumber
    \\
    & = 
    \bra{\Psi} V(0,\Delta t)^{\dagger} O_{\overline{\eta},\Lambda}^{\dagger} \, O_{\eta,\Lambda} V(0,\Delta t) \ket{\Psi} \; ,
\end{align}
where, for convenience, we have defined
\begin{equation}\label{eq:O eta Lambda definition}
    O_{\eta ,\Lambda} := T\left\lbrace \prod_{t\in\mathcal{T}} V(t,t+\Delta t) \prod_{\vec{x}\in\mathcal{X}}\delta(\phi(t,\vec{x}) - \eta(t,\vec{x}))  \right\rbrace \; .
\end{equation}
Now, using the form of the interaction unitary in~\eqref{eq:interaction unitary definition}, we see that for small enough $\Delta t$ we have
\begin{equation}
    V(t,t+\Delta t ) \approx \exp ( -i \Delta t H_{int}^{(I)}(t) ) \; .
\end{equation}
Assuming a $\phi^4$ self interaction, $H_{int}^{(I)}(t)$ is given in~\eqref{eq:interaction unitary definition}. For small spatial lattice spacing, $\Delta x$, $H_{int}^{(I)}(t)$ is approximately
\begin{equation}
    H_{int}^{(I)}(t) \approx - \sum_{\vec{x}\in\mathcal{X}}\Delta x \frac{\lambda}{4!}\, \phi(t,\vec{x})^4 \; ,
\end{equation}
which allows us to write
\begin{align}\label{eq:interaction unitary as product}
    V(t,t+\Delta t ) & \approx  \exp \left( i \Delta t \Delta x \frac{\lambda}{4!} \sum_{\vec{x}\in\mathcal{X}} \phi(t,\vec{x})^4 \right)
    \nonumber
    \\
    &= \prod_{\vec{x}\in\mathcal{X}}e^{ i \Delta V \frac{\lambda}{4!} \phi(t,\vec{x})^4} \; ,
\end{align}
where we have defined the lattice volume element $\Delta V = \Delta t \Delta x$. The last line follows from the fact that, for distinct $\vec{x}$ and $\vec{x}'$, we have $[\phi(t,\vec{x}),\phi(t,\vec{x}')]=0$, and hence $e^{ a\phi(t,\vec{x}) + b \phi(t,\vec{x}') } = e^{ a\phi(t,\vec{x})}e^{b\phi(t,\vec{x}')}$ for constants $a$ and $b$.

Further, we can reorder any of the delta functions in the product $\prod_{\vec{x}\in\mathcal{X}}\delta(\phi(t,\vec{x}) - \eta(t,\vec{x}))$ in~\eqref{eq:O eta Lambda definition}, since they all commute. Replacing $V(t,t+\Delta t )$ in~\eqref{eq:O eta Lambda definition} by the last line of~\eqref{eq:interaction unitary as product}, we can also distribute the terms $e^{ i \Delta V \frac{\lambda}{4!} \phi(t,\vec{x})^4}$ from $V(t,t+\Delta t)$ throughout the product of delta functions in any order we want, since everything commutes. This means we can rewrite
\begin{align}\label{eq:O eta Lambda continuum approx as causal}
    O_{\eta ,\Lambda} & \approx T\left\lbrace \prod_{t\in\mathcal{T}} \prod_{\vec{x}\in\mathcal{X}} e^{ i \Delta V \frac{\lambda}{4!} \phi(t,\vec{x})^4}\delta(\phi(t,\vec{x}) - \eta(t,\vec{x}))  \right\rbrace 
    \nonumber
    \\
    & = C\left\lbrace \prod_{x\in\Lambda} e^{ i \Delta V \frac{\lambda}{4!}\phi(x)^4}\delta(\phi(x) - \eta(x))  \right\rbrace \; ,
\end{align}
where in the last line we replaced the the double product with a single product over lattice points. For the purpose of comparison with certain causal set expressions in Section~\ref{sec:Histories Framework} and~\ref{sec:Introducing an interaction}, we have also replaced the time ordering with a \textit{causal} ordering, denoted by $C\lbrace \cdot\rbrace$. For a product of $r$ operators $O_1(x_1)...O_r(x_r)$, at the spacetime points $x_1,...,x_r$ respectively, $C\lbrace O_1(x_1)...O_r(x_r)\rbrace$ means we order the operators in any way that is consistent with the following rule: if $x_n\prec x_m$, then $O_n(x_n)$ cannot appear to the left of $O_m(x_m)$\footnote{Note that any choice of order satisfying this rule can be related to any other satisfactory choice by the permutation of operators localised at spacelike points. By the commutativity of spacelike operators, this therefore does not change the total operator in question, and hence any satisfactory choice is equivalent.}. We are free to replace the time ordering by a causal ordering in~\eqref{eq:O eta Lambda continuum approx as causal} as any causal ordering can be modified into a time ordering by permuting operators localised at spacelike points, which commute by spacelike commutativity.

If we allow ourselves some more mathematical slack for the time being, we can imagine taking~\eqref{eq:lattice delta functions double path integral} `further' by pinning the configurations $\xi$ and $\overline{\xi}$ to $\eta$ and $\overline{\eta}$ for \textit{every} spacetime point $x\in M$, rather than just for those in our lattice $\Lambda$. The double path integral in the first line of~\eqref{eq:lattice delta functions double path integral} is then no longer an integral, and simply evaluates to $D(\eta , \overline{\eta};\lambda)$ - the decoherence functional evaluated at $\eta$ and $\overline{\eta}$. For convenience, let us relabel $\eta,\overline{\eta}\mapsto\xi,\overline{\xi}$. Now, following~\eqref{eq:lattice delta functions double path integral}, and the last line of~\eqref{eq:O eta Lambda continuum approx as causal}, we are motivated to write down the following formal expression for $D(\xi , \overline{\xi};\lambda)$:
\begin{equation}\label{eq:continuum interacting decoherence functional to canonical framework}
    D(\xi , \overline{\xi} ; \lambda) = \bra{\Psi} C\left\lbrace \prod_{x\in M} e^{ i \, dV \frac{\lambda}{4!}\phi(x)^4}\delta(\phi(x) - \overline{\xi}(x))  \right\rbrace^{\dagger} C\left\lbrace \prod_{x\in M} e^{ i \, dV \frac{\lambda}{4!}\phi(x)^4}\delta(\phi(x) - \xi(x))  \right\rbrace \ket{\Psi} \; ,
\end{equation}
where $dV$ is the infinitesimal volume element. The free theory is given by $\lambda = 0$, and thus for the the free decoherence functional we have the following formal expression:
\begin{equation}\label{eq:continuum free decoherence functional to canonical framework}
    D(\xi , \overline{\xi}) = \bra{\Psi} C\left\lbrace \prod_{x\in M} \delta(\phi(x) - \overline{\xi}(x))  \right\rbrace^{\dagger} C\left\lbrace \prod_{x\in M}\delta(\phi(x) - \xi(x))  \right\rbrace \ket{\Psi} \; .
\end{equation}
We call these expressions `formal' as the products are over all spacetime points $x\in M$ are not well defined in the continuum, but products over all causal set points can be defined, as we will see below. For the sake of comparison with certain causal set expressions in Section~\ref{sec:Histories Framework} and~\ref{sec:Introducing an interaction} (specifically~\eqref{eq:causal set free decoherence functional to canonical framework} and~\eqref{eq:causal set interacting decoherence functional to canonical framework}) it is important to again highlight that the field operators, $\phi(x)$, in these expressions are in the (Heisenberg) interaction picture in the (free) interacting theory.

Recalling the definition of $D(\xi,\overline{\xi};\lambda)$ in equation~\eqref{eq:continuum interacting decoherence functional form}, equation~\eqref{eq:continuum interacting decoherence functional to canonical framework} is then telling us how to write the rhs of~\eqref{eq:continuum interacting decoherence functional form} in terms of the usual quantities of the canonical framework, e.g. field operators and the given state $\ket{\Psi}$. Equation~\eqref{eq:continuum free decoherence functional to canonical framework} is similarly doing this for the free theory. For the causal set case in Sections~\ref{sec:Histories Framework} and~\ref{sec:Introducing an interaction} we will write down an analogous equality (both for the free and interacting theories) between an explicit measure on $\xi$ and $\overline{\xi}$ and expressions in the canonical framework.

Finally, as causal sets are manifestly covariant (at least under the Poisson sprinkling process), we can be sure that our analogous causal set expressions below will not have any residual frame or coordinate dependence. The same cannot be said for the expressions just derived for the regular lattice $\Lambda$.

\section{Causal Set Free Scalar Field Theory}\label{sec:Causal Set Free Scalar Field Theory}

\subsection{Canonical Framework}\label{sec:Canonical Framework}

Here we follow the \textit{Sorkin-Johnston formalism} for real scalar QFT~\cite{sorkin_johnston_state_continuum,stephen_johnston_thesis}. The signature of this formalism is its choice of ground state of the quantum theory. This choice is made using the positive part of the Pauli-Jordan function. We will see precisely what this means shortly. First, let us introduce some basic concepts for causal set QFT.

Consider a fixed causal set $(C,\prec )$ of finite cardinality $N = |C|$ with some \textit{natural labelling}. That is, if $x,y\in C$ are two distinct causal set points with the labels $i,j\in \lbrace 1,...,N\rbrace$ respectively, then if $x\prec y$, then $i<j$. In what follows we will simplify our notation by using $x$ to denote both the causal set point and its label. Thus we will write statements like $x\prec y$, and $x<y$. It should be clear from the context whether the point or the label is being referred to. We will also use the point and its label interchangeably in subscripts/superscripts of matrices and vectors.

We introduce a discreteness length scale, $l_0$, which is assumed to be of order the Planck length. Through this we define a discreteness density $\rho = 1/l_0^d$, where the dimension $d$ must also be chosen for comparison with the continuum theory of the same dimension. In the case that $C$ comes from a sprinkling into a $d$-dimensional spacetime of finite volume, $\rho$ is the sprinkling density.

We define a retarded Green function, $G_{xy}$, where by `retarded' we mean that $G_{xy}=0$ unless $y\prec x$. There are different candidates for the retarded Green function on a causal set. For example, one can lift $G_{xy}$ from the continuum retarded Green function $G(x,y)$~\cite{Sorkin_Yazdi_entropy}. Alternatively, one can define $G_{xy}$ through a `hop-and-stop' model as in~\cite{Johnston_hop_and_stop}. Here we will consider the latter case for concreteness.

With a given choice of retarded Green function we note that the advanced Green function is simply the transpose $G_{xy}^T = G_{yx}$. We then define the Pauli-Jordan matrix $\Delta := G - G^T$, where we have omitted the indices for brevity. It is straightforward to see, then, that the matrix $\frac{1}{\rho}i\Delta$ is Hermitian (the inclusion of $1/\rho$ here allows us to directly compare the eigenvalues with those of the continuum Pauli-Jordan function~\cite{Surya_causet_deSitter_dim_analysis}), and thus we can find a complete orthonormal basis of (complex-valued) eigenvectors and their associated (real-valued) eigenvalues. Here we have assumed the standard inner product on $\mathbb{C}^N$, i.e. $(v,w) = \sum_{x\in C} v^*_x w_x$. 

We further find that if a vector $v$ has a non-zero eigenvalue $\mu$, then the complex conjugate of the vector, $v^*$, has eigenvalue $-\mu$. Thus, we can group eigenvectors based on whether their eigenvalues are positive, negative, or zero. Let $K< N$ be the number of positive eigenvalues, and let $v^{(k)}_x$ be the coefficients of a normalised eigenvector with positive eigenvalue $\mu_k >0$ (where $k=1,...,K$ labels the positive eigenvalues). That is,
\begin{equation}\label{eq:pauli jordan eigenvector}
    \frac{1}{\rho}\sum_{y\in C}i\Delta_{xy}v_y^{(k)} = \mu_k v_x^{(k)} \; .
\end{equation}
The coefficients of the $K$ negative eigenvalue eigenvectors are then $(v_x^{(k)})^*$. The remaining $N-2K$ eigenvectors have eigenvalue $0$, and form a basis of the kernel $\text{ker}\, \Delta$. We can now write
\begin{equation}\label{eq:pauli jordan matrix decomposition}
    i\Delta_{xy} =\rho \sum_{k=1}^K \mu_k \left( v_x^{(k)} (v_y^{(k)})^* - (v_x^{(k)})^* v_y^{(k)}\right) \; .
\end{equation}

Note, by thinking of $\rho^{-1}$ as the discrete volume element $\Delta V$, we see that~\eqref{eq:pauli jordan eigenvector} is analogous to the continuum integral eigenequation
\begin{equation}
    \int_M dV_y \, i \Delta(x,y) v^{(k)}(y) = \mu_k v^{(k)}(x)  \; ,
\end{equation}
for some finite volume spacetime $M$. We can then contrast the causal set eigenvalues directly to their continuum counterparts. Here $dV_y = d^d y \sqrt{-g(y)}$ is the usual spacetime volume element, and $\Delta(x,y)$ is the continuum Pauli-Jordan function, defined in terms of the retarded Green function $G(x,y)$ as $\Delta(x,y)= G(x,y) - G(y,x)$.

Returning to causal set QFT, for each positive eigenvalue eigenvector of $i\Delta$ we introduce annihilation operators $a_k$ (where $k=1,..., K$) satisfying the commutation relations $[a_k , a_l^{\dagger}]=\delta_{kl}\mathds{1}$. The signature step of the Sorkin-Johnston formalism is then to define the normalised ground state, $\ket{\Omega}$, also known as the Sorkin-Johnston state, as that which satisfies $a_k\ket{\Omega} = 0$ for all $k=1,...,K$. We further define orthonormal basis states as $\ket{n_1 , ... , n_K} = \prod_{k=1}^K \frac{(a_k^{\dagger})^{n_k}}{\sqrt{n_k !}}\ket{\Omega}$, where $n_k = 0,1,2,...$. The Fock space of our theory, $\mathcal{F}$, is then the span of these states, and is isomorphic to the (completion of) the tensor product of $K$ harmonic oscillator Hilbert spaces.

For each point $x\in C$ we define the field operator at $x$ as
\begin{equation}\label{eq:causet field expansion in terms of modes}
    \phi_x = \sqrt{\rho}\sum_{k=1}^K \sqrt{\mu_k} \left( v_x^{(k)} a_k + (v_x^{(k)})^* a_k^{\dagger} \right) \; .
\end{equation}
This expansion in terms of modes is comparable to the continuum expansion of the field operator-valued distribution $\phi(x)$ in terms of plane waves, e.g. $\phi(x) = \sum_k u^{(k)}(x) a_k + u^{(k)}(x)^* a_k^{\dagger}$. In the free/interacting continuum theory, expanding $\phi(x)$ in this way, i.e. in terms of modes $u^{(k)}(x)$ that satisfy the free equations of motion, tells us that $\phi(x)$ is in the Heisenberg/Interaction picture, as it carries the free dynamics. Thus, we should think of our causal set field operator, $\phi_x$, as being in the Heisenberg/Interaction picture in the free/interacting causal set QFT.

Using~\eqref{eq:causet field expansion in terms of modes} one can then verify that
\begin{equation}\label{eq:causet field commutation relations}
    [\phi_x , \phi_y] = i\Delta_{xy} \mathds{1} \; ,
\end{equation}
which is analogous to the continuum covariant commutation relations $[\phi(x), \phi(y)] = i\Delta(x,y)$. As in the continuum, if $x,y\in C$ are mutually spacelike, then $\Delta_{xy}=0$, and hence $\phi_x$ commutes with $\phi_y$. In this precise sense we can think of the field operator, $\phi_x$, as `local' to $x$.

In addition, from~\eqref{eq:causet field expansion in terms of modes} we can also determine the 2-point, or Wightman, function of our theory in terms of our modes:
\begin{equation}\label{eq:causet SJ 2 point function}
    W_{xy} = \bra{\Omega}\phi_x \phi_y \ket{\Omega} = \rho \sum_{k=1}^K \mu_k v_x^{(k)} (v_y^{(k)})^* \; .
\end{equation}
Recalling~\eqref{eq:pauli jordan matrix decomposition} we see that this 2-point function is simply the positive part of $i\Delta$.

\vspace{2mm}
\noindent \textbf{Remark} It is worth commenting on the choice of modes and ladder operators implicit in the expansion~\eqref{eq:causet field expansion in terms of modes}, and thus the choice of the (Gaussian) Sorkin-Johnston ground state $\ket{\Omega}$. This choice also led to the specific 2-point function in~\eqref{eq:causet SJ 2 point function}. Alternatively, we could have followed Algebraic QFT and introduced abstract algebra elements, $\phi_x$, for each $x\in C$, satisfying~\eqref{eq:causet field commutation relations}. After this we could then pick any sufficiently well behaved 2-point function $W'$\footnote{By sufficiently well behaved we mean that $W'$ is Hermitian and, given any $v\in\text{ker}\Delta$, $W.v=0$ and $v^{\dagger}.W=0$.}, and use the Gel'fand-Naimark-Segal (GNS) representation theorem to represent the algebra of field operators on some Hilbert space, and in the process define the Gaussian ground state in this Hilbert space as that which has the 2-point function $W'$. This ground state need not coincide with the SJ state $\ket{\Omega}$ above.

\vspace{2mm}
One aspect not yet accounted for is in what sense the causal set field operators satisfy the `equations of motion', which is a standard requirement of the continuum operator-valued distribution $\phi(x)$. In fact, the expansion~\eqref{eq:causet field expansion in terms of modes} takes this into account in a way that is analogous to the continuum case. 

In the continuum the classical solutions we are interested in are those of compact support on Cauchy surfaces. For any such solution, $(\Box + m^2)\varphi(x) = 0$, we can find a test function (a smooth function whose support in spacetime is compact), $f(x)$, such that  
\begin{equation}
    \varphi(x) = \int_M dV_y \Delta(x,y) f(y) \; ,
\end{equation}
We also write this more compactly as $\varphi = \Delta f$. This means that $\text{ker}(\Box + m^2) = \text{im}\Delta$. Given the $L^2$-inner product on complex-valued functions on the spacetime $M$ (where $M$ is of finite volume), $(\varphi ,\xi )_{L^2} = \int_M dV_x \varphi(x)^* \eta(x)$, we then find that \textit{solutions are orthogonal to functions in the kernel of $\Delta$}. 

In the causal set case, we can analogously define a `solution' as a complex-valued function on $C$ that is orthogonal to $\text{ker}\Delta$, or equivalently $\text{ker}\, i\Delta$. Given the expansion of $\phi_x$~\eqref{eq:causet field expansion in terms of modes} in terms of non-zero eigenvalue eigenvectors of $i\Delta$, it is straightforward to verify that, for any $w\in\text{ker}\, i\Delta$, we have $\sum_{x\in C}w_x^* \phi_x = 0$, i.e. $\phi_x$ is orthogonal to $\text{ker}\, i\Delta$ as desired.

\subsection{Histories Framework}\label{sec:Histories Framework}

With the canonical framework in place, we are now ready to define the decoherence functional for a free real scalar field on a fixed causal set, $(C,\prec )$, of finite cardinality $N = |C|$. Recall that we have assumed a natural labelling of $C$. That is, for any pair of distinct points $x,y\in C$, if $x\prec y$ as points, then $x<y$ as labels. 

Now, given the formal continuum expression in~\eqref{eq:continuum free decoherence functional to canonical framework}, we are motivated to define the free decoherence functional as Sorkin did in~\cite{Sorkin_scalar_field_histories}:
\begin{equation}\label{eq:causal set free decoherence functional to canonical framework}
    D(\xi , \overline{\xi}) = \bra{\Omega} \delta(\phi_1 - \overline{\xi}_1) ... \delta(\phi_N - \overline{\xi}_N)\delta(\phi_N - \xi_N)...\delta(\phi_1 - \xi_1)\ket{\Omega} \; ,
\end{equation}
where $\xi$ and $\overline{\xi}$ are real-valued field configurations on $C$, with values $\xi_x$ and $\overline{\xi}_x$ at a given point $x\in C$. 

The ordering of the delta functions according to their natural labelling means we have already implemented the causal ordering of~\eqref{eq:continuum free decoherence functional to canonical framework}. It is worth pointing out that, unlike the uncountably many delta functions in rhs of~\eqref{eq:continuum free decoherence functional to canonical framework}, the finite number of delta functions here pose no problem to the well-definedness of this decoherence functional as a measure over $\mathbb{R}^{2N}$, i.e. over the space of all field configurations $\xi$ and $\overline{\xi}$.

One of the main results of~\cite{Sorkin_scalar_field_histories} is that the rhs of~\eqref{eq:causal set free decoherence functional to canonical framework} can be manipulated into a form analogous to the continuum expression in~\eqref{eq:continuum decoherence functional form}. One finds,
\begin{equation}\label{eq:causal set free decoherence functional form}
    D(\xi , \overline{\xi} ) =\mathcal{N}\, e^{-i\Delta S[\xi,\overline{\xi}]}\, \delta(G^T(\xi - \overline{\xi}))\, e^{-Q[\xi,\overline{\xi}]} \; ,
\end{equation}
where $\Delta S[\xi,\overline{\xi}]$ is a bilinear form in $\xi$ and $\overline{\xi}$, and is analogous to $S[\xi ] - S[\overline{\xi}]$ in the continuum expression~\eqref{eq:continuum decoherence functional form}. $Q[\xi , \overline{\xi}]$ is also a bilinear form, and depends upon the symmetric part of the SJ 2-point function $W$. In this sense, the term $e^{-Q[\xi,\overline{\xi}]}$ takes into account the choice of `initial' state, c.f. the term $\Psi(\overline{\xi}(0))^*\Psi(\xi(0))$ in~\eqref{eq:continuum decoherence functional form}. $\delta(G^T(\xi - \overline{\xi}))$ is a delta function ensuring that $\xi - \overline{\xi}$ is in the kernel of $G^T$ (the advanced Green function). This is analogous to the delta function $\delta ( \xi(T) - \overline{\xi}(T) )$ in~\eqref{eq:continuum decoherence functional form}, which ensures the two field configurations are `pinned together' on the future boundary. The delta function $\delta(G^T(\xi - \overline{\xi}))$, in some sense, does the same job (see~\cite{Sorkin_scalar_field_histories} for further discussion). Finally, there is a normalisation constant, $\mathcal{N}$, at the front to ensure that $\int_{\mathbb{R}^{2N}}d^N\xi d^N \overline{\xi}D(\xi,\overline{\xi}) = 1$. For the explicit forms of $\Delta S$ and $Q$, and for further comparisons to their continuum counterparts, see~\cite{Sorkin_scalar_field_histories}.

One aspect of~\eqref{eq:causal set free decoherence functional form} mentioned in~\cite{Sorkin_scalar_field_histories}, and still unexplored, is how one might recover the continuum expression in~\eqref{eq:continuum decoherence functional form} via some continuum limit ($N\rightarrow\infty$) of~\eqref{eq:causal set free decoherence functional form}.

\section{Causal Set Interacting Scalar Field Theory}\label{sec:Causal Set Interacting Scalar Field Theory}

\subsection{Introducing an interaction}\label{sec:Introducing an interaction}

We now introduce a $\phi^4$ self-interaction to our real scalar field theory. In the continuum this is done by adding an interaction term to the action, $S_{\text{int}}[\xi;\lambda] = \int_M dV\,  \frac{\lambda}{4!}\xi^4$, which, following~\eqref{eq:continuum interacting decoherence functional form}, results in the interacting decoherence functional $D(\xi,\overline{\xi};\lambda) = D(\xi,\overline{\xi})e^{i\frac{\lambda}{4!}\int_M dV \, (\xi^4 - \overline{\xi}^4)}$. In the causal set case, we are then motivated to modify the term $\Delta S[\xi,\overline{\xi}]$ in the free decoherence functional to be
\begin{equation}
    \Delta S[\xi,\overline{\xi};\lambda] = \Delta S[\xi,\overline{\xi}] + \frac{1}{\rho}\sum_{n=1}^N \frac{\lambda_n}{4!} \left(\xi_n^4 - \overline{\xi}_n^4 \right) \; ,
\end{equation}
where we recall that $\frac{1}{\rho}$ is like a discrete volume element $\Delta V$. Here $\lambda\in \mathbb{R}^N$ and has components $\lambda_n$. Allowing $\lambda$ to vary across the causal set $C$ gives us the freedom to describe interactions local to some region of spacetime. 

With this we now have a candidate decoherence functional for an interacting real scalar field theory on a causal set:
\begin{align}\label{eq:causal set interacting decoherence functional form}
    D(\xi , \overline{\xi} ; \lambda ) & = \mathcal{N}(\lambda)\, e^{-i\Delta S[\xi,\overline{\xi};\lambda]}\, \delta(G^T(\xi - \overline{\xi}))\, e^{-Q[\xi,\overline{\xi}]} 
    \nonumber
    \\
    & = \frac{\mathcal{N}(\lambda)}{\mathcal{N}}D(\xi ,\overline{\xi})e^{i\frac{1}{\rho}\sum_{n=1}^N \frac{\lambda_n}{4!} \left(\xi_n^4 - \overline{\xi}_n^4 \right)}\; ,
\end{align}
where the constant $\mathcal{N}(\lambda)$ is chosen so that the decoherence functional is normalised, i.e.
\begin{equation}
    \int_{\mathbb{R}^{2N}}d^N\xi d^N\overline{\xi}\, D(\xi,\overline{\xi};\lambda) = 1 \; .
\end{equation}
It is not clear at first glance whether $\mathcal{N}(\lambda)$ is different than the constant $\mathcal{N}$ in the free case, or whether $\mathcal{N}(\lambda)$ depends on $\lambda$ at all, as indicated by the notation. As mentioned in~\cite{Sorkin_scalar_field_histories}, a further question is whether this new decoherence functional is positive semidefinite, as a general decoherence functional should be. Finally, we were motivated to write down this definition of the interacting decoherence functional by the continuum expression for the interacting decoherence functional in~\eqref{eq:continuum interacting decoherence functional form}, but in the continuum theory one can also (formally) express this interacting decoherence functional via the canonical framework~\eqref{eq:continuum interacting decoherence functional to canonical framework}. There is then the question of whether there exists such a canonical framework representation of the causal set $D(\xi,\overline{\xi};\lambda)$, and further whether this is analogous to the continuum expression in~\eqref{eq:continuum interacting decoherence functional to canonical framework}.

To finish this section let us answer these questions. To begin we rewrite the rhs of~\eqref{eq:causal set interacting decoherence functional form} using the free case definition~\eqref{eq:causal set free decoherence functional to canonical framework}:
\begin{align}\label{eq:rewriting causal set interacting decoherence functional with free case canonical definition}
    D(\xi , \overline{\xi} ; \lambda ) & = \frac{\mathcal{N}(\lambda)}{\mathcal{N}}D(\xi ,\overline{\xi})e^{i\frac{1}{\rho}\sum_{n=1}^N \frac{\lambda_n}{4!} \left(\xi_n^4 - \overline{\xi}_n^4 \right)}
    \nonumber
    \\
    & = \frac{\mathcal{N}(\lambda)}{\mathcal{N}}\bra{\Omega} \delta(\phi_1 - \overline{\xi}_1) ... \delta(\phi_N - \overline{\xi}_N)\delta(\phi_N - \xi_N)...\delta(\phi_1 - \xi_1)\ket{\Omega}e^{i\frac{1}{\rho}\sum_{n=1}^N \frac{\lambda_n}{4!} \left(\xi_n^4 - \overline{\xi}_n^4 \right)} 
    \nonumber
    \\
    & = \frac{\mathcal{N}(\lambda)}{\mathcal{N}}\bra{\Omega} \delta(\phi_1 - \overline{\xi}_1)e^{-i\frac{\lambda_1}{\rho 4!}\overline{\xi}_1^4} ... \delta(\phi_N - \overline{\xi}_N)e^{-i\frac{\lambda_N}{\rho 4!}\overline{\xi}_N^4}
    \nonumber
    \\
    & \hspace{25mm} \times \delta(\phi_N - \xi_N)e^{i\frac{\lambda_N}{\rho 4!}\xi_N^4}...\delta(\phi_1 - \xi_1)e^{i\frac{\lambda_1}{\rho 4!}\xi_1^4}\ket{\Omega} \; .
\end{align}
Now, using the fact that, for any sufficiently well behaved~\footnote{By sufficiently well behaved we mean the following: given the spectral measure $E_i(\cdot )$ associated with $\phi_i$ (which maps Borel subsets of $\mathbb{R}$ to projectors on the Fock space $\mathcal{F}$), and given any state $\ket{\Psi}\in\mathcal{F}$, the function $f:\mathbb{R}\rightarrow\mathbb{C}$ is square integrable against the real measure $\mu(\cdot) = \bra{\Psi}E_i(\cdot) \ket{\Psi}$ (which maps Borel subsets of $\mathbb{R}$ to non-negative real numbers).} function $f:\mathbb{R}\rightarrow\mathbb{C}$, we have
\begin{equation}\label{eq:delta function of phi_i identity}
    \int_{\mathbb{R}}da \, \delta(\phi_i - a)f(a)  = f(\phi_i) = \int_{\mathbb{R}}da \, \delta(\phi_i - a)f(\phi_1) \; ,
\end{equation}
our normalisation condition then gives
\begin{align}\label{eq:interacting normalisation derivation}
    1 & = \int_{\mathbb{R}^{2N}}d^N\xi d^N\overline{\xi} \, D(\xi,\overline{\xi};\lambda) 
    \nonumber
    \\
    & = \frac{\mathcal{N}(\lambda)}{\mathcal{N}}\int_{\mathbb{R}^{2N}}d^N\xi d^N\overline{\xi} \bra{\Omega} \delta(\phi_1 - \overline{\xi}_1)e^{-i\frac{\lambda_1}{\rho 4!}\overline{\xi}_1^4} ... \delta(\phi_N - \overline{\xi}_N)e^{-i\frac{\lambda_N}{\rho 4!}\overline{\xi}_N^4}
    \nonumber
    \\
    & \hspace{50mm} \times \delta(\phi_N - \xi_N)e^{i\frac{\lambda_N}{\rho 4!}\xi_N^4}...\delta(\phi_1 - \xi_1)e^{i\frac{\lambda_1}{\rho 4!}\xi_1^4}\ket{\Omega}
    \nonumber
    \\
    & = \frac{\mathcal{N}(\lambda)}{\mathcal{N}}\bra{\Omega} \int_{\mathbb{R}}d\overline{\xi}_1  \delta(\phi_1 - \overline{\xi}_1)e^{-i\frac{\lambda_1}{\rho 4!}\overline{\xi}_1^4} ... \int_{\mathbb{R}}d\overline{\xi}_N  \delta(\phi_N - \overline{\xi}_N)e^{-i\frac{\lambda_N}{\rho 4!}\overline{\xi}_N^4}
    \nonumber
    \\
    & \hspace{25mm} \times \int_{\mathbb{R}}d\xi_N  \delta(\phi_N - \xi_N)e^{i\frac{\lambda_N}{\rho 4!}\xi_N^4} ... \int_{\mathbb{R}}d\xi_1  \delta(\phi_1 -\xi_1)e^{i\frac{\lambda_1}{\rho 4!}\xi_1^4}\ket{\Omega} 
    \nonumber
    \\
    & = \frac{\mathcal{N}(\lambda)}{\mathcal{N}}\bra{\Omega}e^{-i\frac{\lambda_1}{\rho 4!}\phi_1^4} ... e^{-i\frac{\lambda_N}{\rho 4!}\phi_N^4}e^{i\frac{\lambda_N}{\rho 4!}\phi_N^4}...e^{i\frac{\lambda_1}{\rho 4!}\phi_1^4}\ket{\Omega} \; .
\end{align}
The exponentials then cancel each other, one by one, leaving $1 = \frac{\mathcal{N}(\lambda)}{\mathcal{N}}\braket{\Omega | \Omega}$. Since $\braket{\Omega | \Omega}=1$, we get $\mathcal{N}(\lambda) = \mathcal{N}$, which answers our normalisation question.

Regarding our question about a canonical framework representation of the causal set $D(\xi,\overline{\xi};\lambda)$, analogous to~\eqref{eq:continuum interacting decoherence functional to canonical framework}, equations~\eqref{eq:delta function of phi_i identity} and~\eqref{eq:rewriting causal set interacting decoherence functional with free case canonical definition} tell us that we can write the interacting decoherence functional as
\begin{align}\label{eq:causal set interacting decoherence functional to canonical framework}
    D(\xi,\overline{\xi};\lambda) & = \frac{\mathcal{N}(\lambda)}{\mathcal{N}}\bra{\Omega} \delta(\phi_1 - \overline{\xi}_1)e^{-i\frac{\lambda_1}{\rho 4!}\overline{\xi}_1^4} ... \delta(\phi_N - \overline{\xi}_N)e^{-i\frac{\lambda_N}{\rho 4!}\overline{\xi}_N^4}
    \nonumber
    \\
    & \hspace{25mm} \times \delta(\phi_N - \xi_N)e^{i\frac{\lambda_N}{\rho 4!}\xi_N^4}...\delta(\phi_1 - \xi_1)e^{i\frac{\lambda_1}{\rho 4!}\xi_1^4}\ket{\Omega}
    \nonumber
    \\
    & = \bra{\Omega} \delta(\phi_1 - \overline{\xi}_1)e^{-i\frac{\lambda_1}{\rho 4!}\phi_1^4} ... \delta(\phi_N - \overline{\xi}_N)e^{-i\frac{\lambda_N}{\rho 4!}\phi_N^4}
    \nonumber
    \\
    & \hspace{15mm} \times \delta(\phi_N - \xi_N)e^{i\frac{\lambda_N}{\rho 4!}\phi_N^4}...\delta(\phi_1 - \xi_1)e^{i\frac{\lambda_1}{\rho 4!}\phi_1^4}\ket{\Omega} \; ,
\end{align}
where we have also used our previous result that $\mathcal{N}(\lambda)=\mathcal{N}$. The last expression is immediately comparable to the formal continuum expression in~\eqref{eq:continuum interacting decoherence functional to canonical framework}. The $\rho^{-1}$ factor in each exponent is analogous to the $dV$ volume element in each exponent in~\eqref{eq:continuum interacting decoherence functional to canonical framework}.

The final question of whether $D(\xi,\overline{\xi};\lambda)$ is positive semidefinite is more technically challenging to prove, but likely true nonetheless. Let us sketch an argument here.

To phrase the positive semidefinite property we need to introduce some concepts. The space of possible field configurations $\xi$ is $\mathbb{R}^N$ - one real number for each causal set point. This is our sample space. The event algebra, $\mathfrak{A}$, over this sample space is then the Borel subsets of $\mathbb{R}^N$. Each Borel subset $B\in\mathfrak{A}$, is called an \textit{event}, and corresponds to some subset of field configurations. For any pair of events $B,\overline{B}\in\mathfrak{A}$, we write
\begin{equation}
    D(B,\overline{B};\lambda) = \int_{B}d^N\xi \int_{\overline{B}}d^N\overline{\xi} \, D(\xi,\overline{\xi};\lambda) \; .
\end{equation}
For any finite set of events, $\mathfrak{B}=\lbrace B_1 , ... , B_r \rbrace \subset \mathfrak{A}$, we define the \textit{event matrix} $D_{ab} := D(B_b , B_a ; \lambda)$, for $a,b=1,...,r$. 

Now, we say the decoherence functional is \textit{positive semidefinite} if, for all finite sets of events, the corresponding event matrix is positive semidefinite, i.e. $\sum_{a,b=1}^r v_a^* D_{ab}v_b \geq 0$ for all $v\in\mathbb{C}^r$. This is also called \textit{strong positivity} in~\cite{henry_fay_strong_positivity}.

For any finite set of events $\mathfrak{B}=\lbrace B_1 , ... , B_r \rbrace \subset \mathfrak{A}$, and any $v\in\mathbb{C}^r$, we have
\begin{align}\label{eq:positive semidefinite derivation}
    \sum_{a,b=1}^r v_a^* D_{ab}v_b & = \sum_{a,b=1}^r v_a^* D(B_b , B_a;\lambda) v_b
    \nonumber
    \\
    & = \sum_{a,b=1}^r v_a^* v_b\int_{B_b}d^N\xi \int_{B_a}d^N\overline{\xi} \, D(\xi,\overline{\xi};\lambda)
    \nonumber
    \\
    & = \bra{\Omega} \sum_{a=1}^r v_a^* \int_{B_a}d^N\overline{\xi} \, \delta(\phi_1 - \overline{\xi}_1)e^{-i\frac{\lambda_1}{\rho 4!}\overline{\xi}_1^4} ...  \delta(\phi_N - \overline{\xi}_N)e^{-i\frac{\lambda_N}{\rho 4!}\overline{\xi}_N^4}
    \nonumber
    \\
    & \hspace{10mm} \times \sum_{b=1}^r v_b \int_{B_b}d^N\xi \,  \delta(\phi_N - \xi_N)e^{i\frac{\lambda_N}{\rho 4!}\xi_N^4} ...  \delta(\phi_1 -\xi_1)e^{i\frac{\lambda_1}{\rho 4!}\xi_1^4}\ket{\Omega} \; .
\end{align}
To properly complete this sketch of an argument, one needs to show that 
\begin{equation}\label{eq:well defined operator question}
    O(v,B) := \sum_{b=1}^r v_b \int_{B_b}d^N\xi \,  \delta(\phi_N - \xi_N)e^{i\frac{\lambda_N}{\rho 4!}\xi_N^4} ...  \delta(\phi_1 -\xi_1)e^{i\frac{\lambda_1}{\rho 4!}\xi_1^4} \; ,
\end{equation}
is a well defined operator on the state $\ket{\Omega}$ for any finite set of events $\mathfrak{B}\subset \mathfrak{A}$ and any $v\in\mathbb{C}^r$. Assuming $O(v,B)$ is a well defined operator on $\ket{\Omega}$, then $O(v,B)\ket{\Omega}$ is simply some state $\ket{\Psi}\in\mathcal{F}$, and the last line of~\eqref{eq:positive semidefinite derivation} amounts to $\braket{\Psi | \Psi}$, which is always non-negative.

In fact, to show $O(v,B)$ is a well defined operator on $\ket{\Omega}$, one need only show that ($\ast$) \textit{$O(B)$ is a well defined operator on $\ket{\Omega}$ for any Borel subset $B\subseteq \mathbb{R}^N$, where}
\begin{equation}\label{eq:well defined operator question single Borel}
    O(B) := \int_{B}d^N\xi \,  \delta(\phi_N - \xi_N)e^{i\frac{\lambda_N}{\rho 4!}\xi_N^4} ...  \delta(\phi_1 -\xi_1)e^{i\frac{\lambda_1}{\rho 4!}\xi_1^4} \; .
\end{equation}
If ($\ast$) is the case, then any finite linear combination of operators of the form $O(B)$, as in~\eqref{eq:well defined operator question}, is still a well-defined operator.

Actually proving ($\ast$) is technically involved, and we will not attempt it here. One suggested route for a proof is to start with Borel subsets $B\subseteq \mathbb{R}$ that are $N$-dimensional boxes, i.e. of the form $B = I_1 \times ... \times I_N$ for (potentially unbounded) intervals $I_1 , ... , I_N\subseteq \mathbb{R}$. In this case,
\begin{align}
    O(I_1 \times ... \times I_N ) & = \int_{I_N}d\xi_N \delta(\phi_N - \xi_N)e^{i\frac{\lambda_N}{\rho 4!}\xi_N^4} ... \int_{I_1}d\xi_1 \delta(\phi_1 - \xi_1)e^{i\frac{\lambda_1}{\rho 4!}\xi_1^4}
    \nonumber
    \\
    & = E_N(I_N)e^{i\frac{\lambda_N}{\rho 4!}\phi_N^4}... E_1(I_1)e^{i\frac{\lambda_1}{\rho 4!}\phi_1^4} \; ,
\end{align}
where $E_n(\cdot )$ is the projection-valued measure associated with $\phi_n$, which maps Borel subsets of $\mathbb{R}$ to projectors on the Fock space $\mathcal{F}$. The last line follows from functional calculus~\cite{reed1981functional}. Since the last line is a finite product of projectors and unitaries of the form $e^{i\frac{\lambda_n}{\rho 4!}\phi_n^4}$, it is a well-defined operator.

For a general Borel subset $B\subseteq \mathbb{R}^N$, one can then imagine taking a limit of a disjoint union of boxes that tends to $B$ in some sense, and simultaneously taking the limit of the corresponding operators to furnish a well-defined $O(B)$. Such procedures are routine in measure theory and functional calculus~\cite{reed1981functional}, and will likely not cause any insurmountable roadblocks, only technical challenges. We leave this calculation, and thus the proof of strong positivity, for future work.

\subsection{Interacting 2-point function}\label{sec:Interacting 2-point function}

To make contact with what is done in the continuum, let us focus on the interacting 2-point function $\bra{\Omega}\phi_x^{(H)}\phi_y^{(H)}\ket{\Omega}$, for two causal set points $x,y\in C$. Let us further restrict ourselves to the causally ordered 2-point function, $\bra{\Omega}C\lbrace \phi_x^{(H)}\phi_y^{(H)}\rbrace \ket{\Omega}$, as one can always recover the 2-point function from this if desired. Specifically, if $x\succ y$ or $x=y$ then $\bra{\Omega}\phi_x^{(H)}\phi_y^{(H)}\ket{\Omega}=\bra{\Omega}C\lbrace \phi_x^{(H)}\phi_y^{(H)}\rbrace \ket{\Omega}$, and if $x\prec y$ then $\bra{\Omega}\phi_x^{(H)}\phi_y^{(H)}\ket{\Omega}=\bra{\Omega}C\lbrace \phi_x^{(H)}\phi_y^{(H)}\rbrace \ket{\Omega}^*$.

For convenience, let us then assume that $x\nprec y$, so that $C\lbrace \phi_x^{(H)}\phi_y^{(H)}\rbrace =  \phi_x^{(H)}\phi_y^{(H)}$. Recalling that we have assumed a natural labelling of $C$, we then know that, as labels, $x\geq y$.

To compute $\bra{\Omega}\phi_x^{(H)}\phi_y^{(H)}\ket{\Omega}$ we integrate $\xi_x \xi_y$ against our interacting decoherence functional, $D(\xi,\overline{\xi} ;\lambda)$. Using~\eqref{eq:causal set interacting decoherence functional to canonical framework} we find
\begin{align}\label{eq:causal set interacting 2 point function from decoherence functional}
    \bra{\Omega}\phi_x^{(H)}\phi_y^{(H)}\ket{\Omega} & = \int_{\mathbb{R}^{2N}}d^N \xi d^N \overline{\xi} \, D(\xi,\overline{\xi}; \lambda) \, \xi_x \xi_y
    \nonumber
    \\
    & = \bra{\Omega} V_{x-1,...,1}^{\dagger}\phi_x V_{x-1,...,y} \phi_y V_{y-1,...,1}\ket{\Omega} \; ,
\end{align}
where we have defined the unitary operator
\begin{equation}
    V_{b,b-1,...,a+1,a} := e^{i\frac{\lambda_b}{\rho 4!}\phi_b^4}e^{i\frac{\lambda_{b-1}}{\rho 4!}\phi_{b-1}^4} ... e^{i\frac{\lambda_{a+1}}{\rho 4!}\phi_{a+1}^4}e^{i\frac{\lambda_a}{\rho 4!}\phi_a^4} \; ,
\end{equation}
for two labels $a<b$. By inserting $\mathds{1} = V_{y-1,...,1} V_{y-1,...,1}^{\dagger}$ we can further rewrite the last line of~\eqref{eq:causal set interacting 2 point function from decoherence functional} as
\begin{align}
    \bra{\Omega}\phi_x^{(H)}\phi_y^{(H)}\ket{\Omega} & = \bra{\Omega} V_{x-1,...,1}^{\dagger}\phi_x V_{x-1,...,y} \phi_y V_{y-1,...,1}\ket{\Omega}
    \nonumber
    \\
    & = \bra{\Omega} V_{x-1,...,1}^{\dagger}\phi_x V_{x-1,...,y} ( V_{y-1,...,1} V_{y-1,...,1}^{\dagger} ) \phi_y V_{y-1,...,1}\ket{\Omega} 
    \nonumber
    \\
    & = \bra{\Omega} V_{x-1,...,1}^{\dagger}\phi_x V_{x-1,...,1} V_{y-1,...,1}^{\dagger} \phi_y V_{y-1,...,1}\ket{\Omega} \; .
\end{align}
The last line reveals the unitary relationship between the interacting picture fields, $\phi_x$, which carry the free dynamics, and the Heisenberg fields, $\phi_x^{(H)}$, which carry the full dynamics. Concretely, we have $\phi_x^{(H)} = V_{x-1,...,1}^{\dagger}\phi_x V_{x-1,...,1}$ for any $x\in C$.

We can simplify this unitary relationship slightly. Since all of the exponentials, $e^{i\frac{\lambda_z}{\rho 4!}\phi_z^4}$, in $V_{x-1,...,1}$ are in terms of fields $\phi_z$ for which the labels satisfy $z < x$, we know that the corresponding points $z\in C$ are either to the past of $x$ or spacelike to $x$. If we further assume a natural labelling for which all of the points to the past of $x$ have smaller labels than those that are spacelike to $x$, then the unitary splits into the product $V_{x-1,...,1} = V_{x-1,...,r+1}V_{r,...,1}$, where $r$ is the number of points to the past of $x$. Since all of the exponentials in $V_{x-1,...,r+1}$ are in terms of fields which commute with $\phi_x$, we know that $V_{x-1,...,r+1}^{\dagger}\phi_x V_{x-1,...,r+1} = \phi_x$, and hence $V_{x-1,...,1}^{\dagger}\phi_x V_{x-1,...,1}$ reduces to an expression involving on the unitaries $e^{i\frac{\lambda_z}{\rho 4!}\phi_z^4}$ for which $z\prec x$, i.e.
\begin{equation}\label{eq:unitary relationship between interaction and heisenberg picture fields}
    \phi_x^{(H)} = V_x^{\dagger}\phi_x V_x \; ,
\end{equation}
where $V_x$ is the product of unitary operators $e^{i\frac{\lambda_z}{\rho 4!}\phi_z^4}$ (for $z\prec x$), arranged in any order consistent with the causal ordering on the points $z$ in the past of $x$.

The finiteness of the causal set $C$ means that, after expanding the exponentials in~\eqref{eq:unitary relationship between interaction and heisenberg picture fields}, $\phi_x^{(H)}$ amounts to a finite series in $\lambda$ and the (interaction picture) fields $\phi_z$ for which $z\prec x$. To see this, consider the adjoint action of some exponential $e^{i\frac{\lambda_z}{\rho 4!}\phi_z^4}$ on $\phi_x$, i.e. $e^{-i\frac{\lambda_z}{\rho 4!}\phi_z^4}\phi_x e^{i\frac{\lambda_z}{\rho 4!}\phi_z^4}$. Using the formula
\begin{equation}\label{eq:adjoint exponential action expansion}
    e^{-B}A e^B = \sum_{n=0}^{\infty} \frac{1}{n!}[A , B]_n \; ,
\end{equation}
where $[A,B]_0 = A$, and $[A,B]_{n+1} = [[A,B]_n , B]$, we find
\begin{align}\label{eq:adjoint exponential action on field}
    e^{-i\frac{\lambda_z}{\rho 4!}\phi_z^4}\phi_x e^{i\frac{\lambda_z}{\rho 4!}\phi_z^4} & = \sum_{n=0}^{\infty}\frac{1}{n!}\left( i\frac{\lambda_z}{\rho 4!} \right)^n [\phi_x , \phi_z^4]_n 
    \nonumber
    \\
    & = \phi_x + i\frac{\lambda_z}{\rho 4!} [\phi_x , \phi_z^4]
    \nonumber
    \\
    & = \phi_x - \frac{\lambda_z}{\rho 3!} \Delta_{xz} \phi_z^3 \; ,
\end{align}
 where line 2 follows as all higher commutators with $\phi_z^4$ vanish. Now, if then consider the adjoint action of $e^{i\frac{\lambda_w}{\rho 4!}\phi_w^4}$ on the last line, we get
 \begin{align}
     & e^{-i\frac{\lambda_w}{\rho 4!}\phi_w^4} \phi_x e^{i\frac{\lambda_w}{\rho 4!}\phi_w^4} - \frac{\lambda_z}{\rho 3!} \Delta_{xz} e^{-i\frac{\lambda_w}{\rho 4!}\phi_w^4}\phi_z^3e^{i\frac{\lambda_w}{\rho 4!}\phi_w^4}
     \nonumber
     \\
     = & e^{-i\frac{\lambda_w}{\rho 4!}\phi_w^4} \phi_x e^{i\frac{\lambda_w}{\rho 4!}\phi_w^4} - \frac{\lambda_z}{\rho 3!} \Delta_{xz} \left( e^{-i\frac{\lambda_w}{\rho 4!}\phi_w^4}\phi_z e^{i\frac{\lambda_w}{\rho 4!}\phi_w^4}\right)^3 
     \nonumber
     \\
     = & \phi_x - \frac{\lambda_w}{\rho 3!} \Delta_{xw} \phi_w^3 - \frac{\lambda_z}{\rho 3!} \Delta_{xz} \left( \phi_z - \frac{\lambda_w}{\rho 3!} \Delta_{zw} \phi_w^3 \right)^3 \; ,
 \end{align}
 which makes it clear that each adjoint action amounts to replacing each field as $\phi_x \mapsto \phi_x - \frac{\lambda_y}{\rho 3!} \Delta_{xy} \phi_y^3$. Since $C$ is finite, we only have finitely many adjoint actions, and thus $\phi_x^{(H)}$ is a finite series in $\lambda$ and the (interaction picture) fields to the past of $x$.

Going further, we can actually invert this relationship to write any interaction picture field $\phi_x$ in terms of Heisenberg picture fields $\phi_z^{(H)}$ for $z$ to the past of $x$. This can be done in a recursive fashion starting at the minimal elements of $C$. A minimal element $x\in C$ has nothing to its past, and thus $\phi_x^{(H)} = \phi_x$. Moving on to elements $x$ with only minimal elements to their past, we then know that $\phi_x^{(H)}$ is simply $\phi_x$ plus some cubic terms in any fields $\phi_y$ for which $y$ is minimal and to the past of $x$. Such a $\phi_y$ can be written as $\phi_y^{(H)}$, since $y$ is minimal. We can then express $\phi_x^{(H)} = \phi_x + O\left({\phi_y}^3\right) = \phi_x + O\left({\phi_y^{(H)}}^3\right)$, which can be rearranged to give $\phi_x$ purely in terms of Heisenberg picture fields. This recursive argument can then be continued until all interaction picture fields are expressed in terms of Heisenberg picture fields.

With this relationship in hand, we can define the algebra of the interacting theory abstractly, i.e. without reference to a Hilbert space representation. Given any two Heisenberg fields $\phi_x^{(H)}$ and $\phi_y^{(H)}$, we can determine their commutator, $[\phi_x^{(H)} , \phi_y^{(H)}]$, purely in terms of other Heisenberg fields. To do this, we first expand each Heisenberg field in terms of the interaction picture fields. We then use the commutation relations for interaction picture fields, i.e. $[\phi_z , \phi_w] = i\Delta_{zw}\mathds{1}$, to write $[\phi_x^{(H)} , \phi_y^{(H)}]$ as a finite polynomial in interaction picture fields. Finally, we use our inverse relationship to rewrite each such interaction picture field back in terms of Heisenberg fields. The result will likely be some complicated expression in terms of Heisenberg fields to the past of $x$ and $y$, but it will be a finite series in $\lambda$, and can be computed in principle. Together with the, already defined, action of $\dagger$ on polynomials in Heisenberg fields, these commutation relations written purely in terms of Heisenberg fields define the abstract $*$-algebra for the theory. 

Before moving on, let us compare $\bra{\Omega}\phi^{(H)}_x \phi^{(H)}_y\ket{\Omega}$ with what is computed in the continuum, i.e. equation~\eqref{eq:peskin path integral}. Writing down the ratio of the two path integrals on rhs of~\eqref{eq:peskin path integral} in the causal set case \footnote{Note we have double path integrals in the causal set case. To remedy this we simply integrate out the $\overline{\xi}$ field first. The resulting expression is then comparable to~\eqref{eq:peskin path integral}.} results in the following expression in the canonical framework:  
\begin{equation}\label{eq:causal set analogue of continuum 2-point function}
     \frac{\int_{\mathbb{R}^{2N}}d^N \xi d^N \overline{\xi} \, D(\xi , \overline{\xi}; \lambda ) \, \xi_x \xi_y}{\int_{\mathbb{R}^{2N}}d^N \xi d^N \overline{\xi} \, D(\xi , \overline{\xi}; \lambda )}= \frac{\bra{\Omega}C\lbrace \phi_x \phi_y V_x \rbrace\ket{\Omega}}{\bra{\Omega}C\lbrace V_x \rbrace\ket{\Omega}} \; .
\end{equation}
It is unclear whether the rhs is the correct object to compute, as there is no analogue in the causal set case for the limit $T\rightarrow \infty (1-i\epsilon)$ that is taken in the continuum. We can make the causal set larger, i.e. $N\rightarrow \infty$, but this could only ever be comparable to $T\rightarrow \infty$ in the continuum. We also note that by expanding~\eqref{eq:causal set analogue of continuum 2-point function} in powers of $\lambda$, one sees that the series does not terminate, unlike the finite series of $\bra{\Omega}\phi^{(H)}_x \phi^{(H)}_y\ket{\Omega}$. Let us leave the question of which object is the correct one for future investigations, and continue on with our computation of $\bra{\Omega}\phi^{(H)}_x \phi^{(H)}_y\ket{\Omega}$.

\subsection{The Analogue of Feynman diagrams}\label{sec:The Analogue of Feynman diagrams}

\subsubsection{Introduction}

\begin{figure}
    \centering
    \subfigure[]{\label{fig:phi41loop}\includegraphics[scale=0.23]{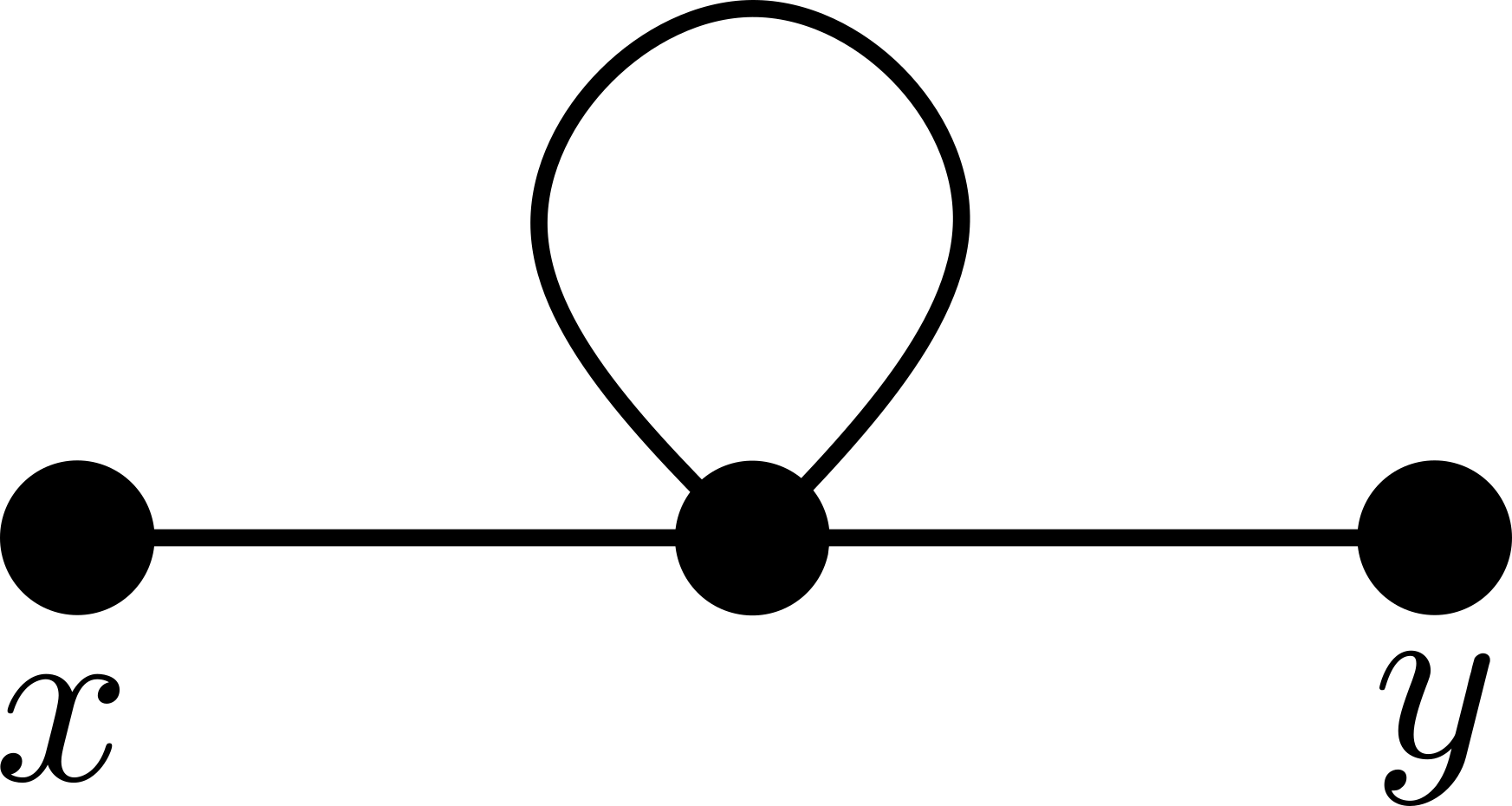}}\hspace{20mm}
    \subfigure[]{\label{fig:phi42loops}\includegraphics[scale=0.23]{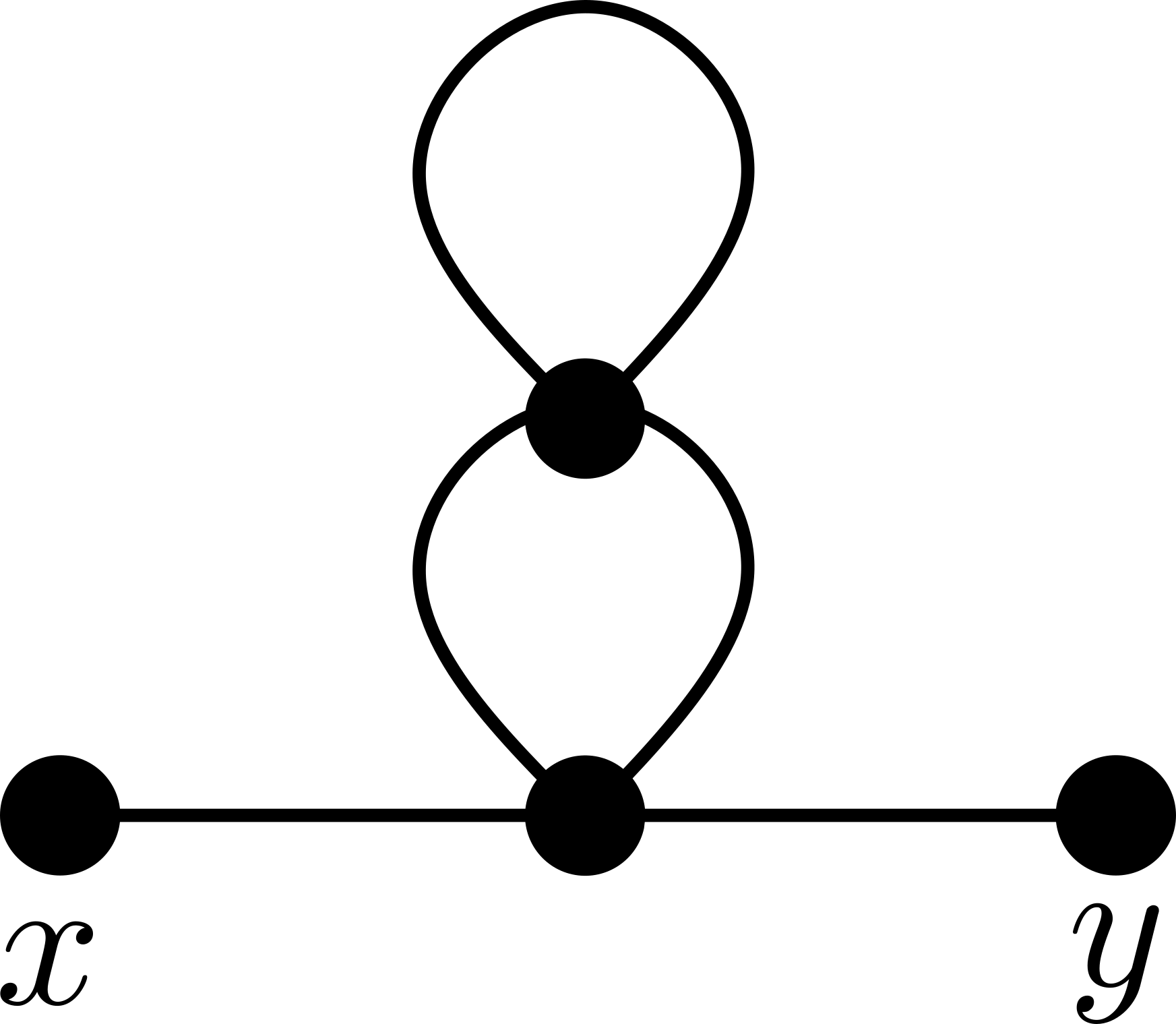}}\hspace{20mm}
    \subfigure[]{\label{fig:phi4circle}\includegraphics[scale=0.23]{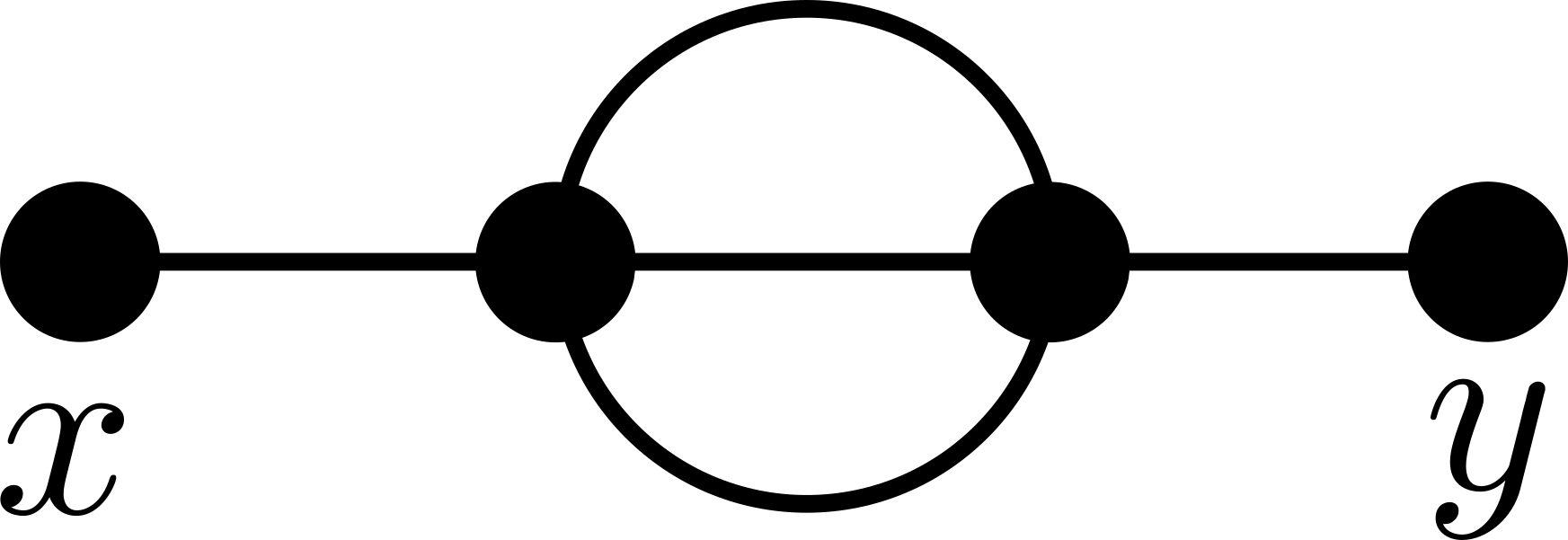}}
    \caption{\footnotesize{Examples of connected diagrams in the 2-point function of a $\phi^4$ theory.}}
    \label{fig:phi4 example diagrams}
\end{figure}

Here we work towards a diagram-based algorithm, akin to Feynman diagrams, for computing $\bra{\Omega}\phi_x^{(H)}\phi_y^{(H)}\ket{\Omega}$ order by order in $\lambda$. The diagrams that arise in our case are almost identical to the usual $\phi^4$ diagrams that appear in the perturbative expansion of the 2-point function (see Figure~\ref{fig:phi4 example diagrams}, or Chapter 4 in~\cite{peskin1995introduction} for examples), but are more complicated in that they contain two types of lines, or edges, between vertices. There are normal, or undirected edges, and directed edges indicated with arrows.

As a reminder, in continuum $\phi^4$ theory~\cite{peskin1995introduction} each vertex, $z$, in a diagram contributes an integral $(-i\lambda)\int_M d^d z$, and each edge between two vertices $x$ and $y$ contributes a factor of the \textit{Feynman propagator} (time-ordered 2-point function) $G^F(x,y) = \bra{\Omega}C\lbrace \phi^{(H)}(x) \phi^{(H)}(y) \ket{\Omega}$. Here we have replaced the time ordering by a causal ordering, as the two are equivalent (since, if the two points are spacelike, spacelike commutativity allows us to reorder them however we please). Finally, one must also divide by the symmetry factor of the diagram.

Returning to the causal set case, let us assume for simplicity that $x\nprec y$, and thus, as labels, $x\geq y$. We further assume a natural labelling for which all points not to the past of $x$ and $y$ have larger labels than $x$ (and hence also $y$). Thus, for any point, $z$, to the past of $x$ or $y$, we know that its label satisfies $z<x$. Recalling~\eqref{eq:causal set interacting 2 point function from decoherence functional}, we have
\begin{align}\label{eq:product of H phix and phiy}
    \phi_x^{(H)}\phi_y^{(H)} & = V_{x-1,...,1}^{\dagger}\phi_x V_{x-1,...y}\phi_y V_{y-1,...,1} 
    \nonumber 
    \\
    & = V_{y-1,...,1}^{\dagger}\left( V_{x-1,...y}^{\dagger}\phi_x V_{x-1,...y} \right) \phi_y V_{y-1,...,1} \; .
\end{align}
For the term in brackets we use the first line of~\eqref{eq:adjoint exponential action on field} to get
\begin{align}
    & V_{x-1,...y}^{\dagger}\phi_x V_{x-1,...y} 
    \nonumber
    \\
    = & \sum_{n_{y},...,n_{x-1}=0}^{\infty}\frac{1}{n_y ! ... n_{x-1}!} \left(\frac{i}{\rho 4!}\right)^{n_y + ... + n_{x-1}} \lambda_{x-1}^{n_{x-1}}...\lambda_y^{n_y}[...[\phi_x , \phi_{x-1}^4]_{n_{x-1}},...,\phi_y^4]_{n_y} \; .
\end{align}
It should then be clear that the last line of~\eqref{eq:product of H phix and phiy} can be written as 
\begin{align}
    V_{y-1,...,1}^{\dagger} \left( V_{x-1,...y}^{\dagger}\phi_x V_{x-1,...y} \right) & \phi_y V_{y-1,...,1}
    \nonumber
    \\
    = & \sum_{n_1,...,n_{x-1}=0}^{\infty}\frac{1}{n_1 ! ... n_{x-1}!} \left(\frac{i}{\rho 4!}\right)^{n_1 + ... + n_{x-1}} \lambda_{x-1}^{n_{x-1}}...\lambda_1^{n_1}
    \nonumber
    \\
    & \hspace{5mm} \times [...[[...[\phi_x , \phi_{x-1}^4]_{n_{x-1}},...,\phi_y^4]_{n_y}\times \phi_y , \phi_{y-1}^4]_{n_{y-1}},...,\phi_1^4]_{n_1} \; .
\end{align}
To tidy this up we introduce the notation $[\shortleftarrow , A]$ to mean $B[\shortleftarrow , A] = [B,A]$, and we follow the convention that composition goes to the right, e.g. $A[\shortleftarrow , B][\shortleftarrow , C] = [A,B][\shortleftarrow , C] = [[A,B],C]$. Now we can rewrite the previous equation as
\begin{align}\label{eq:sum with ns and label ordering}
    V_{y-1,...,1}^{\dagger} \left( V_{x-1,...y}^{\dagger}\phi_x V_{x-1,...y} \right) & \phi_y V_{y-1,...,1}
    \nonumber
    \\
    = & \sum_{n_1,...,n_{x-1}=0}^{\infty}\frac{1}{n_1 ! ... n_{x-1}!} \left(\frac{i}{\rho 4!}\right)^{n_1 + ... + n_{x-1}} \lambda_{x-1}^{n_{x-1}}...\lambda_1^{n_1}
    \nonumber
    \\
    & \hspace{25mm} \times L \lbrace \phi_x \phi_y [\shortleftarrow, \phi_{x-1}^4]_{n_{x-1}}... [\shortleftarrow ,\phi_1^4]_{n_1} \rbrace \; ,
\end{align}
where we have also introduced the notation $L\lbrace \cdot \rbrace$ to denote the \textit{labelled order}, which orders both field operators, $\phi_z$, and commutators, $[\shortleftarrow , \phi_z^4]$, as if they were both the same abstract object, e.g. $L\lbrace \phi_2 \phi_4[\shortleftarrow , \phi_1^4] [\shortleftarrow , \phi_3^4]  \rbrace = \phi_4   [\shortleftarrow , \phi_3^4] \phi_2 [\shortleftarrow , \phi_1^4]$, which then evaluates to $[[\phi_4,\phi_3^4]\phi_2,\phi_1^4]$ under our composition convention. We can rewrite~\eqref{eq:sum with ns and label ordering} as an expansion in orders of $\lambda$ as
\begin{align}\label{eq:expansion of phi x phi y in s}
    V_{y-1,...,1}^{\dagger} \left( V_{x-1,...y}^{\dagger}\phi_x V_{x-1,...y} \right) & \phi_y V_{y-1,...,1}
    \nonumber
    \\
    = & \sum_{n=0}^{\infty}\frac{1}{n!}\left(\frac{i}{\rho 4!}\right)^n\sum_{z_1 , ... ,z_n = 1}^{x-1}  \lambda_{z_1}...\lambda_{z_n}L\lbrace \phi_x \phi_y [\shortleftarrow , \phi_{z_1}^4]...[\shortleftarrow , \phi_{z_n}^4] \rbrace \; ,
\end{align}
which is the first step towards determining all possible Feynman diagrams that contribute towards $\bra{\Omega}\phi_x^{(H)}\phi_y^{(H)}\ket{\Omega}$. More specifically, our diagrams must include two vertices for $x$ and $y$, with 1 free leg each. Each term in the above sum then corresponds to a possible number of \textit{internal vertices} to add, i.e. one vertex for each $z_i$ ($i=1,...,n$). Each internal vertex comes with 4 free legs, corresponding to the power of 4 that each $\phi_{z_i}$ appears with (see Figure~\ref{fig:step1} for example).

\subsubsection{Pre-diagrams}

\begin{figure}
    \centering
    \subfigure[\scriptsize{Initial setup with the vertices in order.}]{\label{fig:step1}\includegraphics[scale=0.23]{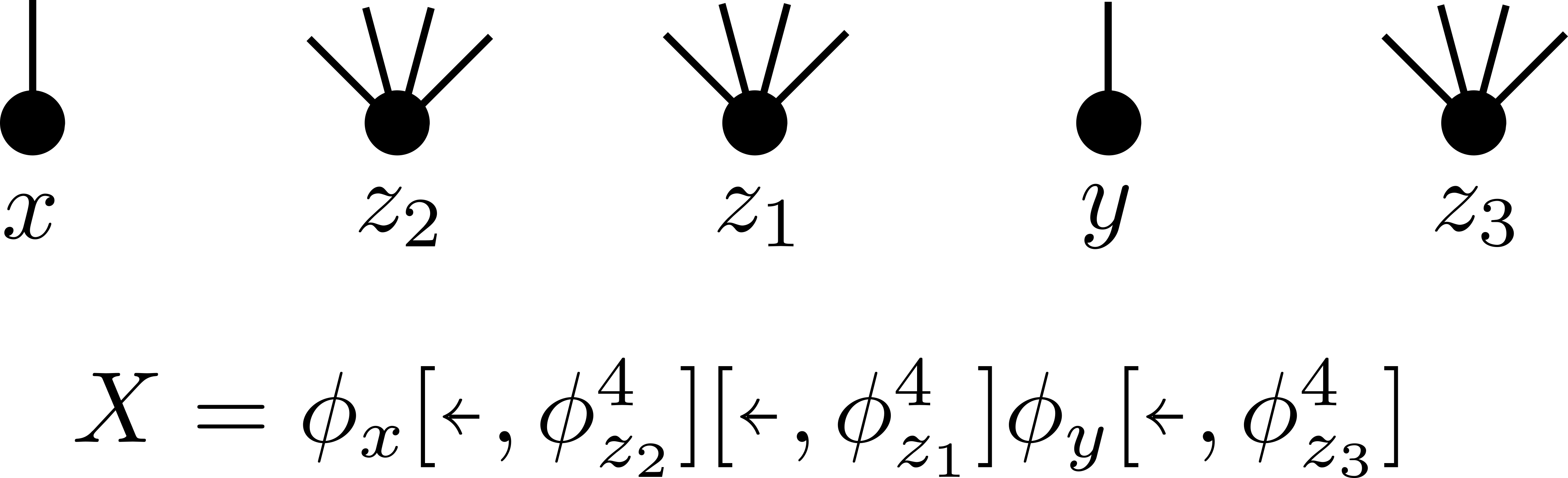}}\hspace{20mm}
    \subfigure[\scriptsize{Factor of $(4)_2$: 4 ways to pick one of $z_2$'s four free legs to join with $x$.}]{\label{fig:step2}\includegraphics[scale=0.23]{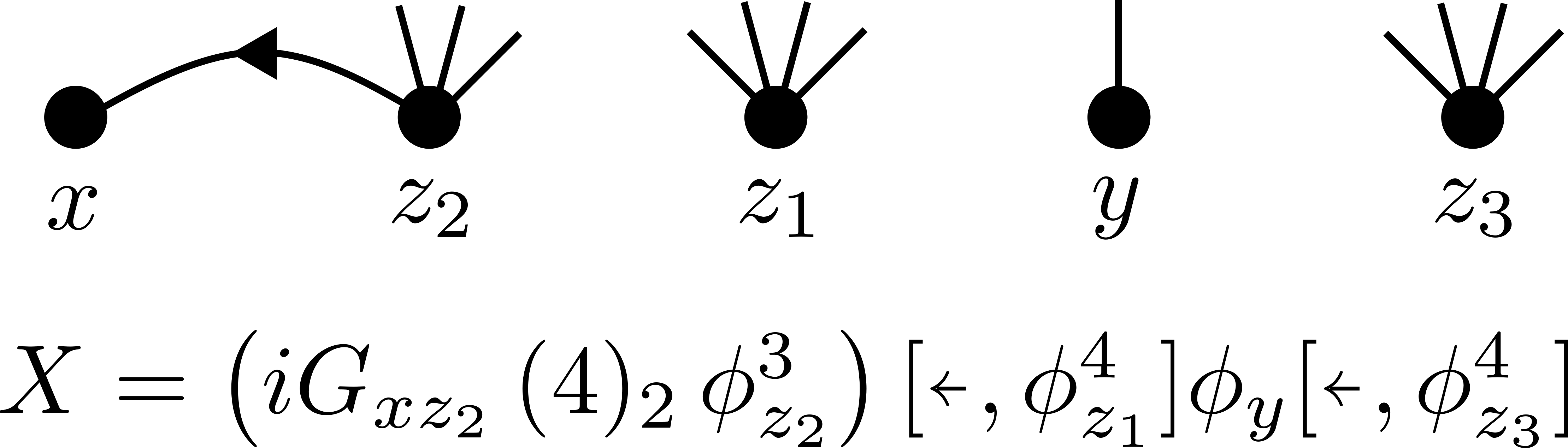}}\\
    \subfigure[\scriptsize{Factor of $(-4.3)_1$: 4 ways to pick the first, then 3 ways to pick the second of $z_1$'s free legs to join with $z_2$. Factor of $(3)_2$: $3=\binom{3}{2}$ ways to pick two of $z_2$'s three free legs to join with the two chosen legs from $z_1$.}]{\label{fig:step3}\includegraphics[scale=0.23]{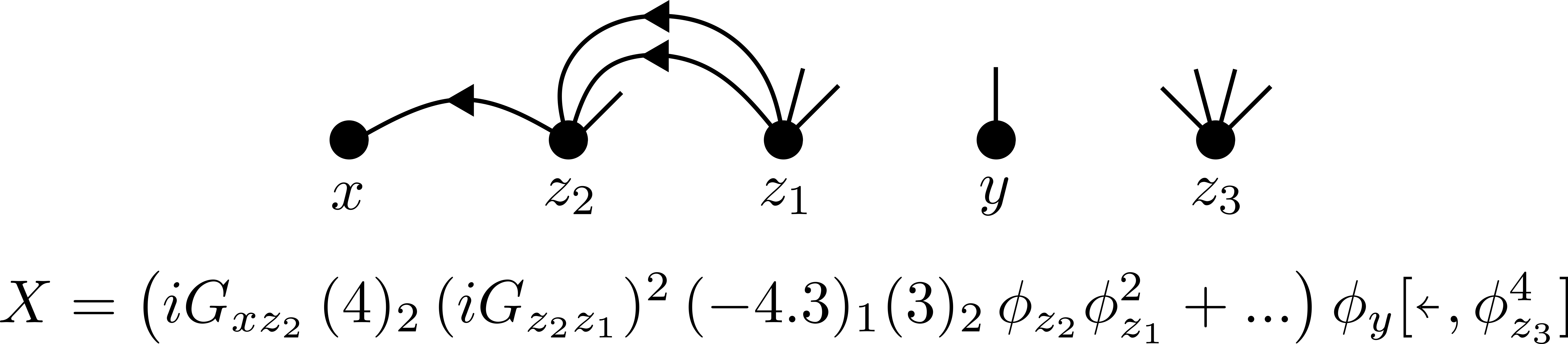}}\\
    \subfigure[\scriptsize{Factor of $(4)_3$: 4 ways to pick one of $z_3$'s four free legs to join with $y$.}]{\label{fig:step4}\includegraphics[scale=0.23]{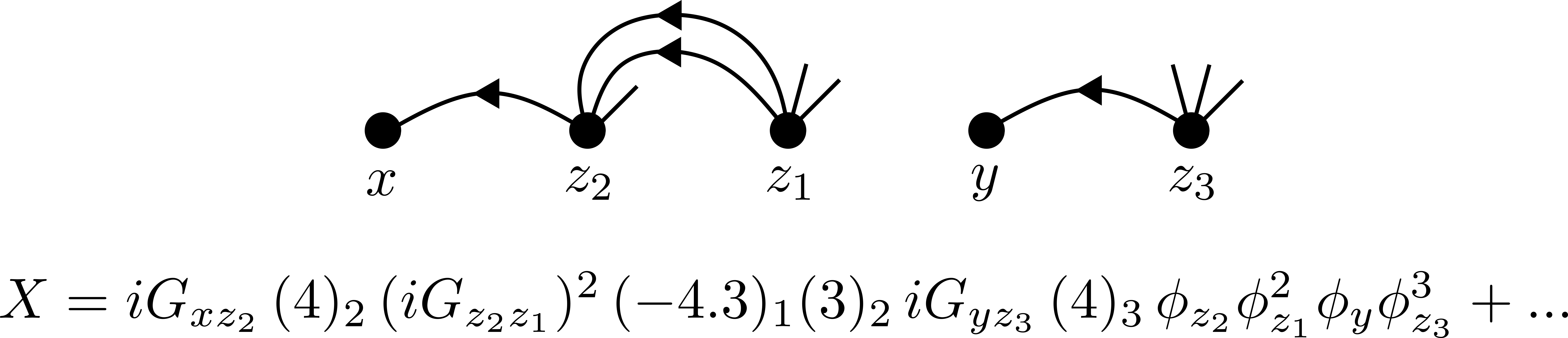}}\\
    \subfigure[\scriptsize{Factor of $(3)_{23}$: 3 ways to pick one of $z_3$'s three remaining legs to join with $z_2$. Factor of $(2)_{13}$: 2 ways to pair up the remaining two legs from $z_1$ and $z_3$. The rhs shows the same labelled diagram but arranged like a Feynman diagram in the continuum theory.}]{\label{fig:finalstep}\includegraphics[scale=0.25]{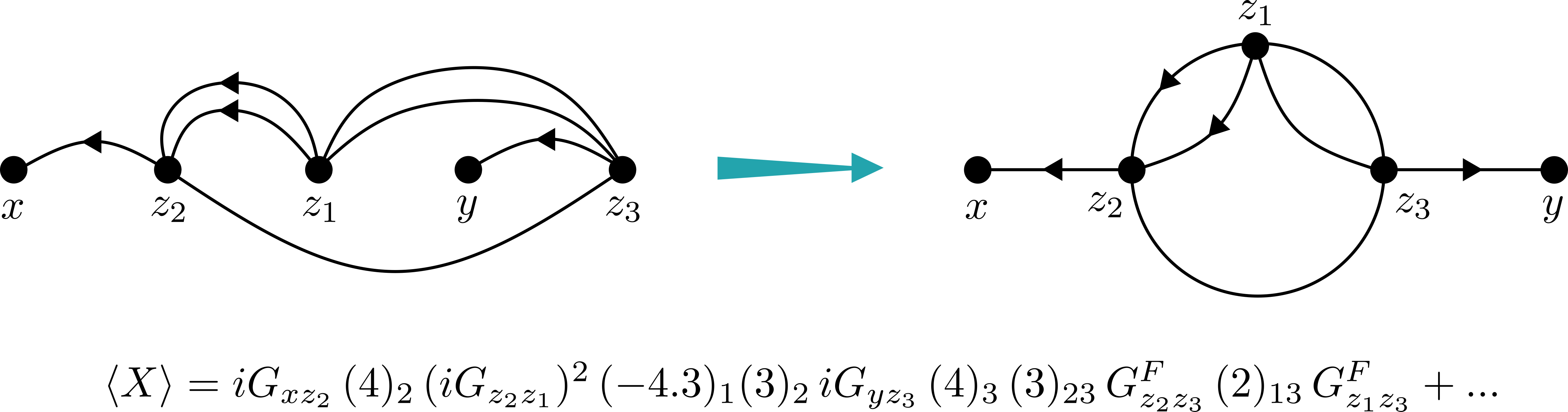}}\\
    \caption{\footnotesize{Step-by-step construction of a \textit{pre-diagram} ((a) to (d)), and then a \textit{labelled diagram} in (e). Underneath each graph we have kept track of the relevant term in $X=\phi_x[\shortleftarrow , \phi_{z_2}^4][\shortleftarrow , \phi_{z_1}^4]\phi_y [\shortleftarrow , \phi_{z_3}^3]$, and $\langle X \rangle \equiv \bra{\Omega} X \ket{\Omega}$ in (e). The subscripts on any numerical factors, e.g. the ``$1$'' in $(-4.3)_1$, indicate the vertex, or vertices, that contributed to the given factors.}}
    \label{fig:diagram contruction}
\end{figure}

Consider some fixed $n$ and some particular values of $z_1,...,z_n$ in the second sum $\sum_{z_1,...,z_n=1}^{x-1}$ in~\eqref{eq:expansion of phi x phi y in s}. We now craft what we call \textit{pre-diagrams}. To do this we first lay out the vertices $x,y,z_1,...,z_n$ from left to right, going from largest to smallest. For example, if $n=3$, and if the labels satisfy the chain of inequalities $x>z_2 > z_1 > y \geq z_3$, then we have the order of vertices shown in Figure~\ref{fig:step1}. Laying out the vertices in this way takes into account the label ordering, $L\lbrace \cdot \rbrace$, in~\eqref{eq:expansion of phi x phi y in s}. Clearly as we go through the different values of $z_1,...,z_n$ in the second sum $\sum_{z_1,...,z_n=1}^{x-1}$, this order will differ. Let us stick with the example in Figure~\ref{fig:diagram contruction} for now.

To make a pre-diagram we then work through the internal vertices - the $z_i$'s - moving from left to right, i.e. in our example we do $z_2$ first (Figure~\ref{fig:step2}), then $z_1$ (Figure~\ref{fig:step3}), and then $z_3$ (Figure~\ref{fig:step4}). For each $z_i$ we join 1 to 4 of its legs to any of the free legs coming from the vertices \textit{to the left} of the given $z_i$ (as in Figure~\ref{fig:diagram contruction}). The resulting edges are all directed to the left, as indicated by the arrows on the lines. Once we have worked through all the internal vertices in order, we call the result a pre-diagram - so-called because not all the legs are paired up, e.g. Figure~\ref{fig:step4}.

Focusing on a given $z_i$ in this process, the act of joining up some non-zero number of its legs encodes the action of $[\shortleftarrow , \phi_{z_i}^4]$ in the expression $\phi_x [\shortleftarrow , \phi_{z_2}^4][\shortleftarrow , \phi_{z_1}^4]\phi_y [\shortleftarrow , \phi_{z_3}^4]$ (note how the order of these operations matches that in Figure~\ref{fig:diagram contruction}). To see this, consider some points whose labels are ordered as $a_n \geq ...\geq a_1 \geq a_0$. It is not too hard to show the identity
\begin{align}\label{eq:commutator identity power of 4}
    & [\phi_{a_n} ... \phi_{a_1} , \phi_{a_0}^4 ]
    \nonumber
    \\
    & = \left(\sum_{r=1}^{n} i\Delta_{a_r a_0}\phi_{a_n} ... (\mathds{1})_{a_r} ... \phi_{a_1} \right)4\phi_{a_0}^3
    \nonumber
    \\
    &  + \left( \sum_{r>r'} i\Delta_{a_r a_0} i\Delta_{a_{r'} a_0} \phi_{a_n} ... (\mathds{1})_{a_r} ... (\mathds{1})_{a_{r'}} ... \phi_{a_1} \right)(-4.3)\phi_{a_0}^2
    \nonumber
    \\
    & + \left( \sum_{r>r'>r''} i\Delta_{a_r a_0}i\Delta_{a_{r'} a_0}i\Delta_{a_{r''} a_0}\phi_{a_n} ... (\mathds{1})_{a_r} ... (\mathds{1})_{a_{r'}} ... (\mathds{1})_{a_{r''}} ... \phi_{a_1} \right) 4.3.2 \phi_{a_0}
    \nonumber
    \\
    & + \left( \sum_{r>r'>r''>r'''} i\Delta_{a_r a_0}i\Delta_{a_{r'} a_0}i\Delta_{a_{r''} a_0}i\Delta_{a_{r'''} a_0}\phi_{a_n} ... (\mathds{1})_{a_r} ... (\mathds{1})_{a_{r'}} ... (\mathds{1})_{a_{r''}} ... (\mathds{1})_{a_{r'''}} ... \phi_{a_1} \right)
    \nonumber
    \\
    & \hspace{5mm} \times (-4.3.2.1) \, ,
\end{align}
where $(X)_{a_r}$ means we insert the operator $X$ instead of $\phi_{a_r}$ in the $r$'th position in the product $\phi_{a_n} ... \phi_{a_1}$. Note that all of the products of fields on the rhs are label ordered. It is also clear that lines 1 to 4 on the rhs can be interpreted as joining 1 to 4 of $a_0$'s 4 free legs with the free legs corresponding to each $a_i$ (for $i=1,...,n$). Each such join comes with a factor of $i\Delta$, which, because the indices of $\Delta$ are always label ordered, can actually be replaced by $iG$, where $G$ is the retarded Green function (this explains the factors of $iG$ in Figure~\ref{fig:diagram contruction}). We also get a factor of $-1$ if we join up an even number of $a_0$'s legs (as in Figure~\ref{fig:step4}). The factors of $4$, $3$, and $2$ come from the number of different choices we have when picking from $a_0$'s legs. Thus, when constructing a pre-diagram, it helps to think of any free legs as being labelled.

Back to our example expression $\phi_x [\shortleftarrow , \phi_{z_2}^4][\shortleftarrow , \phi_{z_1}^4]\phi_y [\shortleftarrow , \phi_{z_3}^4]$. We can now express this as a sum of all possible pre-diagrams constructed from the vertices $x,z_2,z_1,y,z_3$, in that order. For a given pre-diagram, such as that in Figure~\ref{fig:step4}, we can read off the corresponding operator expression as follows. For every directed edge from a vertex $a$ to $b$, we get a factor $iG_{ba}$. If a vertex has $p$ directed edges coming out of it (where $p=1,...,4$), then we get a factor of $(-1)^{p-1}$. For each internal vertex we get a factor of $4!/(4-q)!$, where $q$ is the number of ingoing plus outgoing directed edges for the vertex. If between two vertices there are $s$ directed edges, we get a factor of $1/s!$. Finally, for each free leg coming from a vertex $a$ we get a field operator $\phi_a$. All the field operators must be ordered according to the order of the vertices. Following these rules for the pre-diagram in Figure~\ref{fig:step4} gives
\begin{equation}\label{eq:pre diagram example term}
    - (4.3.2)(4.3)(4)(1/2!) \, iG_{x z_2} \, (i G_{z_2 z_1})^2 \, iG_{y z_3} \, \phi_{z_2}\phi_{z_1}^2 \phi_{z_3}^3 \; ,
\end{equation}
where $(-1)$ comes from the two directed legs leaving $z_1$, $(4.3.2)$ comes from the 3 ingoing plus outgoing directed legs from $z_2$, $(4.3)$ from the 2 outgoing directed legs from $z_1$, $(4)$ from the 1 outgoing leg from $z_1$, and $(1/2!)$ from the 2 directed legs between $z_1$ and $z_2$. The powers of the final fields represent how many free legs remain for each vertex.

\subsubsection{Labelled diagrams}

In the end, we are interested in computing the expectation value of $\phi^{(H)}_x\phi^{(H)}_y$ using the state $\ket{\Omega}$. Thus, we must take the expectation value of the operator expressions arising from any pre-diagrams. The Gaussian nature of $\ket{\Omega}$ means that, when acting on the product of fields $\phi_{z_2}\phi_{z_1}^2\phi_{z_3}^3 =\phi_{z_2}\phi_{z_1}\phi_{z_1}\phi_{z_3}\phi_{z_3}\phi_{z_3} $, for example, we get a sum over all the possible ways to pair up the fields into free 2-point functions. Some of these pairings result in the same expression in terms of 2-point functions, which introduces numerical factors into the result. For our example we get
\begin{equation}\label{eq:wick example}
    \bra{\Omega}\phi_{z_2}\phi_{z_1}\phi_{z_1}\phi_{z_3}\phi_{z_3}\phi_{z_3}\ket{\Omega} = 6 W_{21}W_{13}W_{33} + 3 W_{23}W_{11}W_{33} + 6 W_{23}{W_{13}}^2 \; ,
\end{equation}
where the numerical factors count the possible ways to pair up the fields to get the associated 2-point functions. Note that the indices of the free 2-point functions are all ordered in a way that is consistent with the ordering of the fields in the original product. Since our fields were originally ordered based on their natural labelling, this means that any 2-point functions that appear can be written as causally ordered 2-point functions, or equivalently, Feynman propagators, $G^F_{ab} := \bra{\Omega}C\lbrace \phi_a \phi_b \rbrace \ket{\Omega}$. This explains the factors of $G^F$ in Figure~\ref{fig:finalstep}.

For us, the different terms on the rhs of~\eqref{eq:wick example} correspond to all the possible ways to pair up the remaining legs in our pre-diagram in Figure~\ref{fig:step4}. The 3rd term on the rhs of~\eqref{eq:wick example} corresponds to the pairing shown in Figure~\ref{fig:finalstep}. Pairing up the free legs of a pre-diagram furnishes a \textit{labelled diagram} - so-called because its vertices are labelled $z_1,z_2,$ and so on. In a labelled diagram every internal vertex, $z_i$, will have at least one route via directed edges to get to either $x$ or $y$. Furthermore, every such route one can take (as there may be multiple choices of directed edges one can take from a given internal vertex) leads to either $x$ or $y$ (there are no closed loops of directed edges). The factors for a given labelled diagram are the same as above but with a factor of $G^F_{ab}$ for every undirected edge between any two vertices $a$ and $b$. There may also be a numerical factor if the same labelled diagram can arise in multiple ways from the same pre-diagram. The free legs of the pre-diagram in Figure~\ref{fig:step4} can be paired in 6 different ways to get the labelled diagram in Figure~\ref{fig:finalstep} - $3$ ways to pick one of $z_3$'s three free legs to pair with $z_2$, then 2 ways to pair the remaining legs from $z_1$ and $z_3$. 

To summarise, we now have
\begin{align}\label{eq:2 point function with sum over completed diagrams}
    \bra{\Omega}\phi^{(H)}_x \phi^{(H)}_y \ket{\Omega} = \sum_{n=0}\frac{1}{n!}\left( \frac{i}{\rho 4!}\right)^n \sum_{z_1 ,... , z_n =1}^{x-1} \lambda_{z_1}...\lambda_{z_n} \left[ \begin{array}{l}\text{all possible labelled diagrams} \\ \text{given the order of $z_i$ vertices} \end{array}\right] \, ,
\end{align}
where by ``all possible labelled diagrams ...", we mean all those constructed by putting the vertices in their label order and following the procedure outlined above, i.e. first constructing the pre-diagram by forming directed edges, going one vertex at a time from left to right, then by joining up remaining legs as undirected edges. For different values of $z_1,...,z_n$ in the sum $\sum_{z_1,...,z_n=1}^{x-1}$, the possible labelled diagrams that can arise differ. Each labelled diagram contributes factors of $iG$ for any directed edges, factors of the Feynman propagator $G^F$ for any undirected edges, some numerical factors, and potentially a factor of $-1$.

Consider the labelled diagram in Figure~\ref{fig:finalstep} which contributes the term shown under the graphs in Figure~\ref{fig:finalstep}. We can make this look more like a typical diagram from the continuum $\phi^4$ theory by rearranging the vertices, as we do on the rhs of Figure~\ref{fig:finalstep}. We note, however, that our labelled diagrams differ from those in the continuum in that ours have both directed and undirected edges, and, at this stage, they have labelled vertices. We address the latter point now.

\subsubsection{Unlabelled diagrams}

\begin{figure}
    \centering
    \subfigure[\scriptsize{Pre-diagram to labelled diagram with starting order $x\geq z_1 \geq y \geq z_2$.}]{\label{fig:z1yz2order}\includegraphics[scale=0.23]{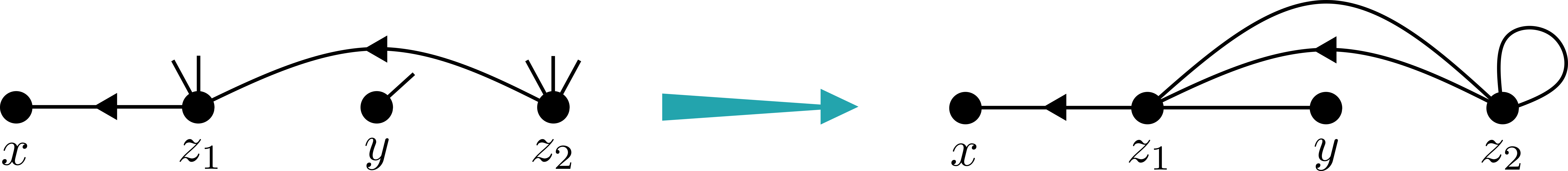}}\hspace{20mm}
    \subfigure[\scriptsize{Pre-diagram to labelled diagram with starting order $x\geq y > z_1 \geq z_2$.}]{\label{fig:yz1z2order}\includegraphics[scale=0.23]{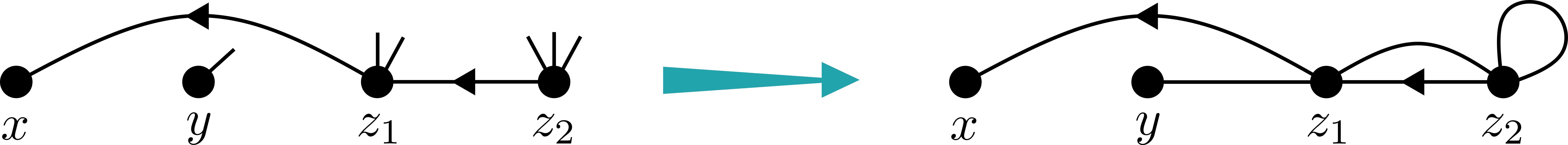}}\\
    \subfigure[\scriptsize{Resulting labelled diagram}]{\label{fig:labelleddiagramz1z2example}\includegraphics[scale=0.23]{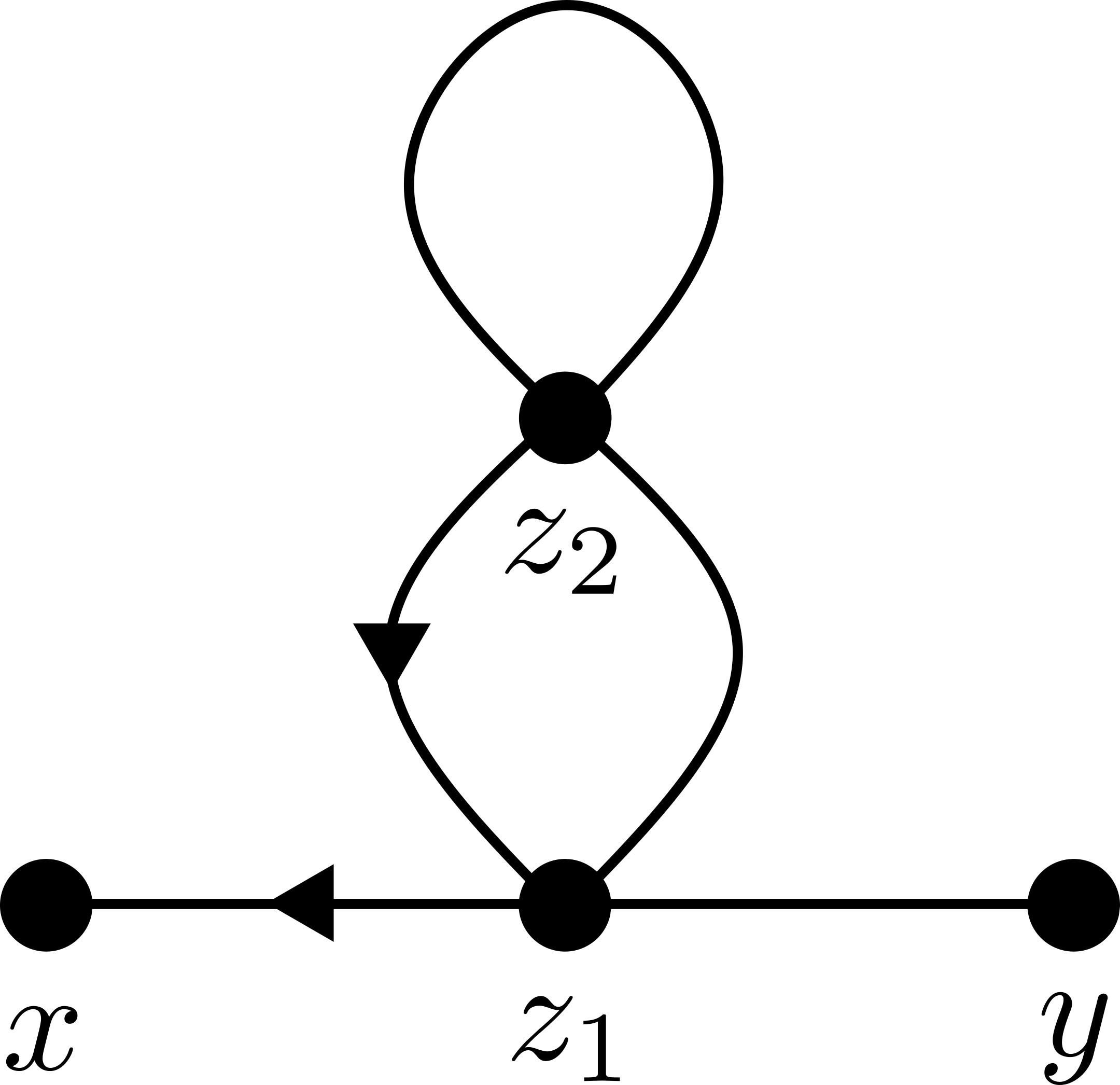}}\hspace{20mm}
    \subfigure[\scriptsize{Invalid pre-diagram for starting order $z_2> z_1$.}]{\label{fig:invalidprediagram}\includegraphics[scale=0.23]{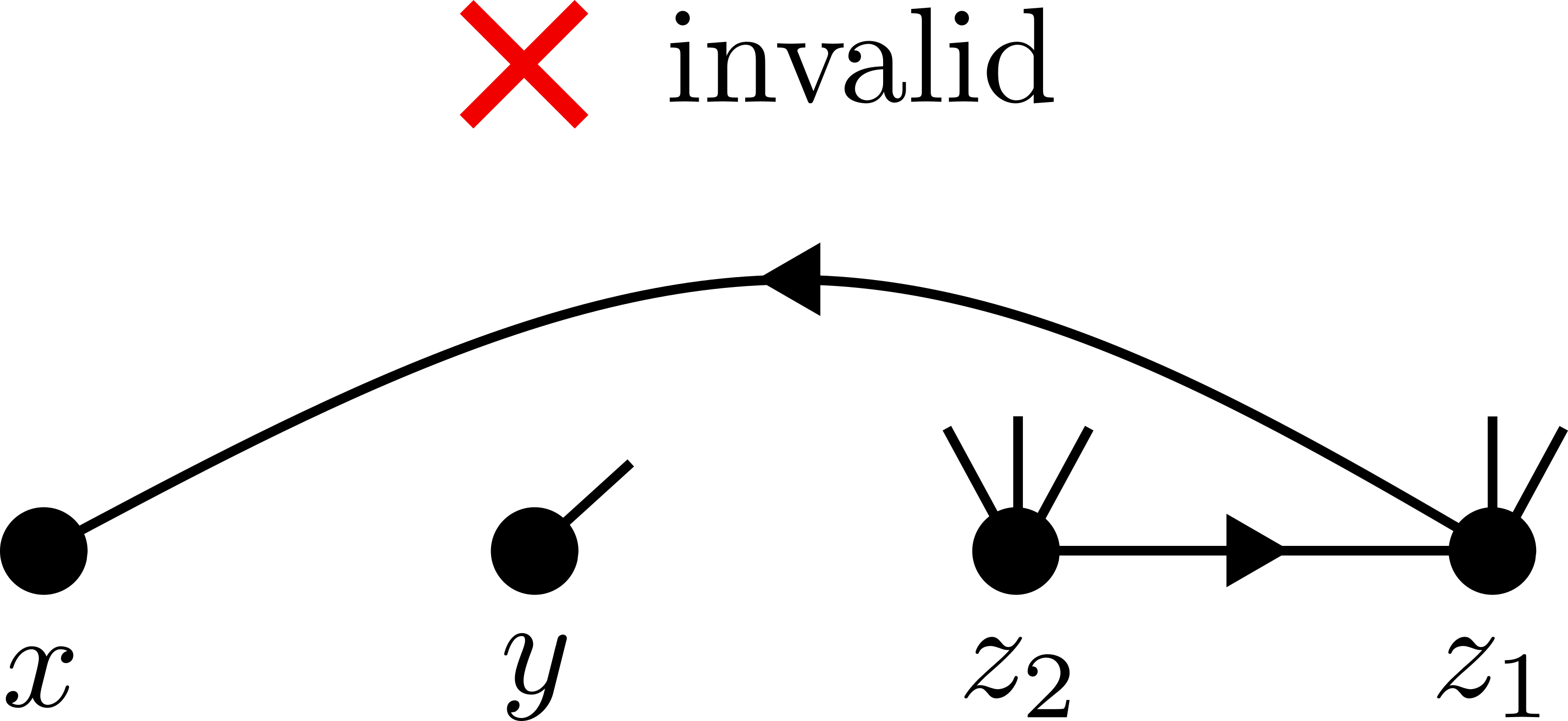}}\hspace{20mm}
    \caption{\footnotesize{Example of how different starting orders can give the same pre-diagram, and hence the same labelled diagram. (c) shows the labelled diagram that results from both (a) and (b). (d) is an invalid pre-diagram as a directed edge goes to the right.}}
    \label{fig:diagram order example}
\end{figure}

Consider the labelled diagram in Figure~\ref{fig:labelleddiagramz1z2example}. This arises as a possible labelled diagram when the values of $z_1$ and $z_2$ in the sum $\sum_{z_1,z_2=1}^{x-1}$ are ordered as $x\geq z_1 \geq y \geq z_2$ (Figure~\ref{fig:z1yz2order}), or as $x\geq y > z_1 \geq z_2$ (Figure~\ref{fig:yz1z2order}). Thus, the ordering between $z_1$ and $y$ does not matter, which one can infer from the undirected edge between $z_1$ and $y$ in the labelled diagram (the rhs of Figures~\ref{fig:z1yz2order} and~\ref{fig:yz1z2order}, or Figure~\ref{fig:labelleddiagramz1z2example}).

If we change the ordering between any vertices connected by a directed edge, however, e.g. if we consider values in the sum for which $z_2>z_1$, then this labelled diagram does not arise (see the invalid pre-diagram in Figure~\ref{fig:invalidprediagram}). It is more convenient, however, to think of it as being there, but since it comes with a factor of $G_{z_1 z_2}$ (which vanishes for $z_2>z_1$), its contribution vanishes.

In general, then, we can think of every $(n+2)$-vertex labelled diagram as appearing in \textit{every term} of the sum $\sum_{z_1,...,z_n=1}^{x-1}$, and only for those values of $z_1,...,z_n$ which match the ordering implied by the directed edges of the labelled diagram does the given diagram contribute something non-trivial.

For any labelled diagram, we also get diagrams corresponding to perturbations of its vertex labels, e.g. swapping $z_1$ and $z_2$ in Figure~\ref{fig:labelleddiagramz1z2example}, or $z_1$, $z_2$ and $z_3$ in Figure~\ref{fig:finalstep}. This suggests that we group together all labelled diagrams that are related via a perturbation of their vertices into one \textit{unlabelled diagram}. This takes into account the factor of $1/n!$ in~\eqref{eq:2 point function with sum over completed diagrams}. With this we can also `absorb' the sum $\left(\frac{i}{\rho}\right)^n \sum_{z_1,...,z_n=1}^{x-1}$ into these new unlabelled diagrams by letting each vertex, $z$, of the unlabelled diagram contribute a sum $\frac{i}{\rho}\sum_{z=1}^{x-1}$.

Recall that the range of $z$ in this sum ensures that the point $z$ is always to the past of the points $x$ or $y$ in the causal set, since we are using a natural labelling for which all points not to past of $x$ and $y$ have larger labels than $x$ (and hence also $y$). We can actually increase the range of this sum to $\sum_{z\in C}$, as in any diagram we know that the vertex $z$ is connected via some route of directed edges to the vertices $x$ or $y$, and hence there will be factors of $G$ that are only non-trivial if the point $z$ is to the past of the points $x$ or $y$. Finally, the factors of $1/4!$ in~\eqref{eq:2 point function with sum over completed diagrams} will cancel any numerical factors arising in the above procedure, up to the remaining symmetry factor of the unlabelled diagram. Henceforth, we call such an unlabelled diagram simply a \textit{diagram}.

\subsubsection{Summary of analogue Feynman diagrams and rules}\label{sec:Summary of analogue Feynman diagrams and rules}

Following the previous section we have
\begin{align}
    \bra{\Omega}\phi^{(H)}_x \phi^{(H)}_y\ket{\Omega} = \diagram{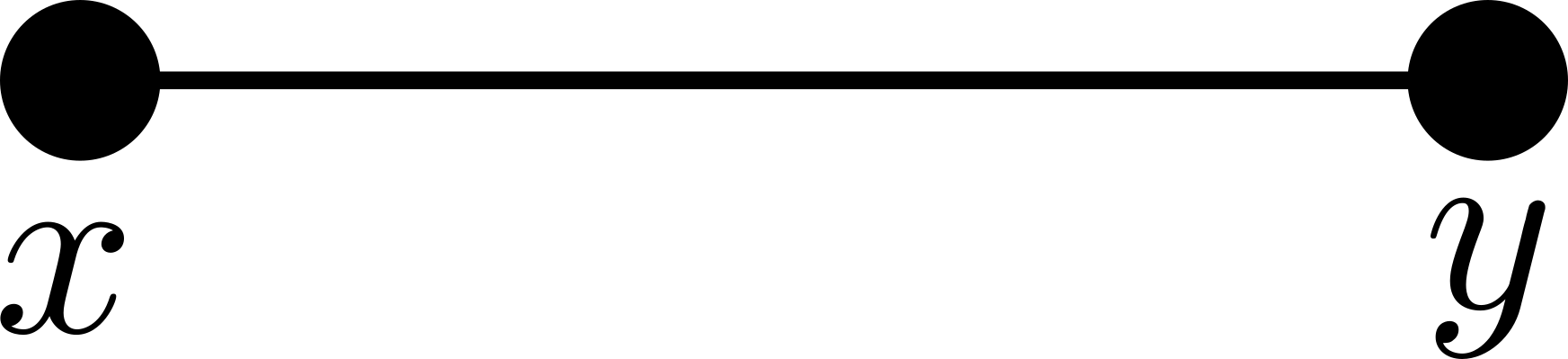} + \diagram{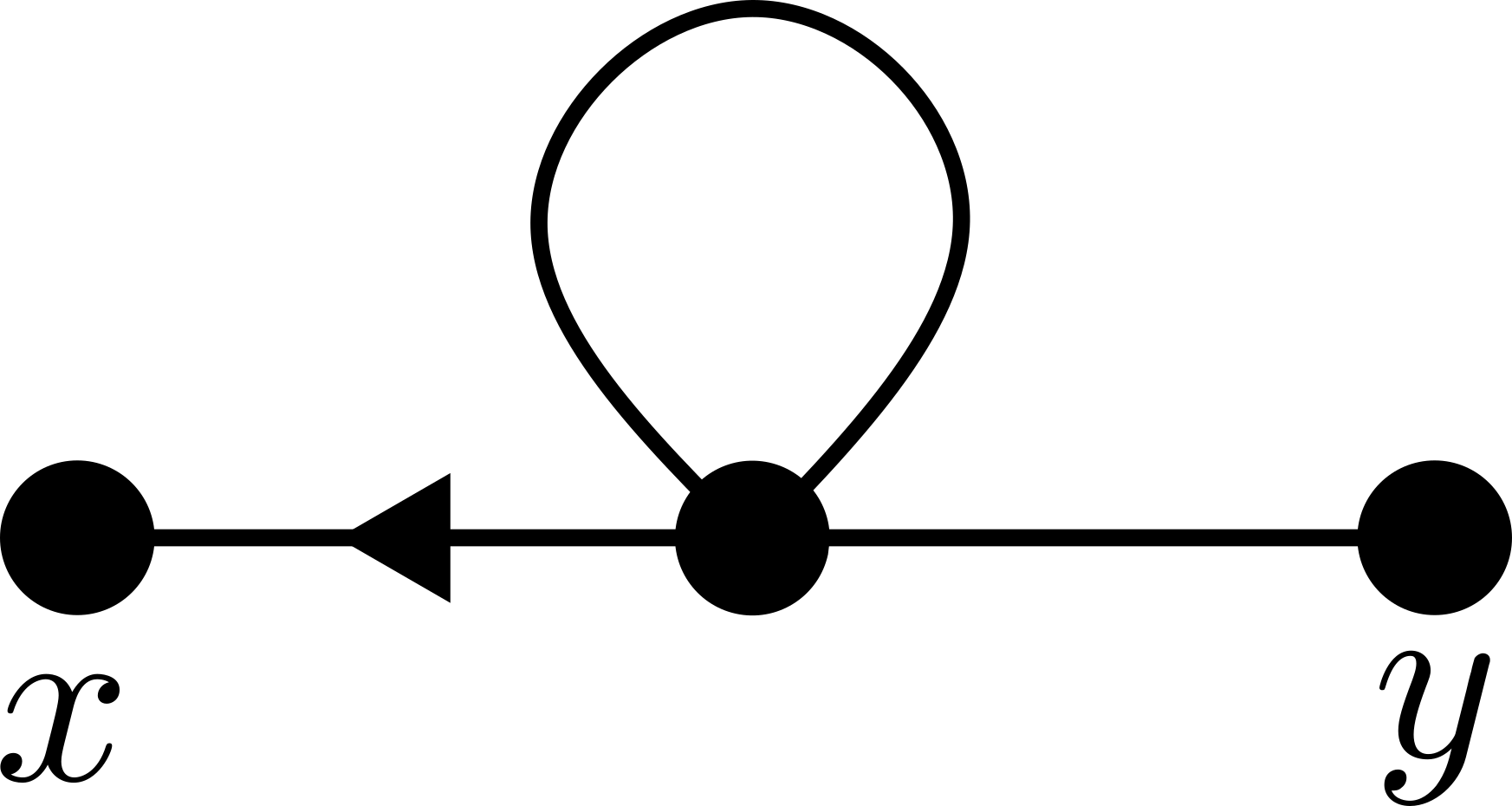} + \diagram{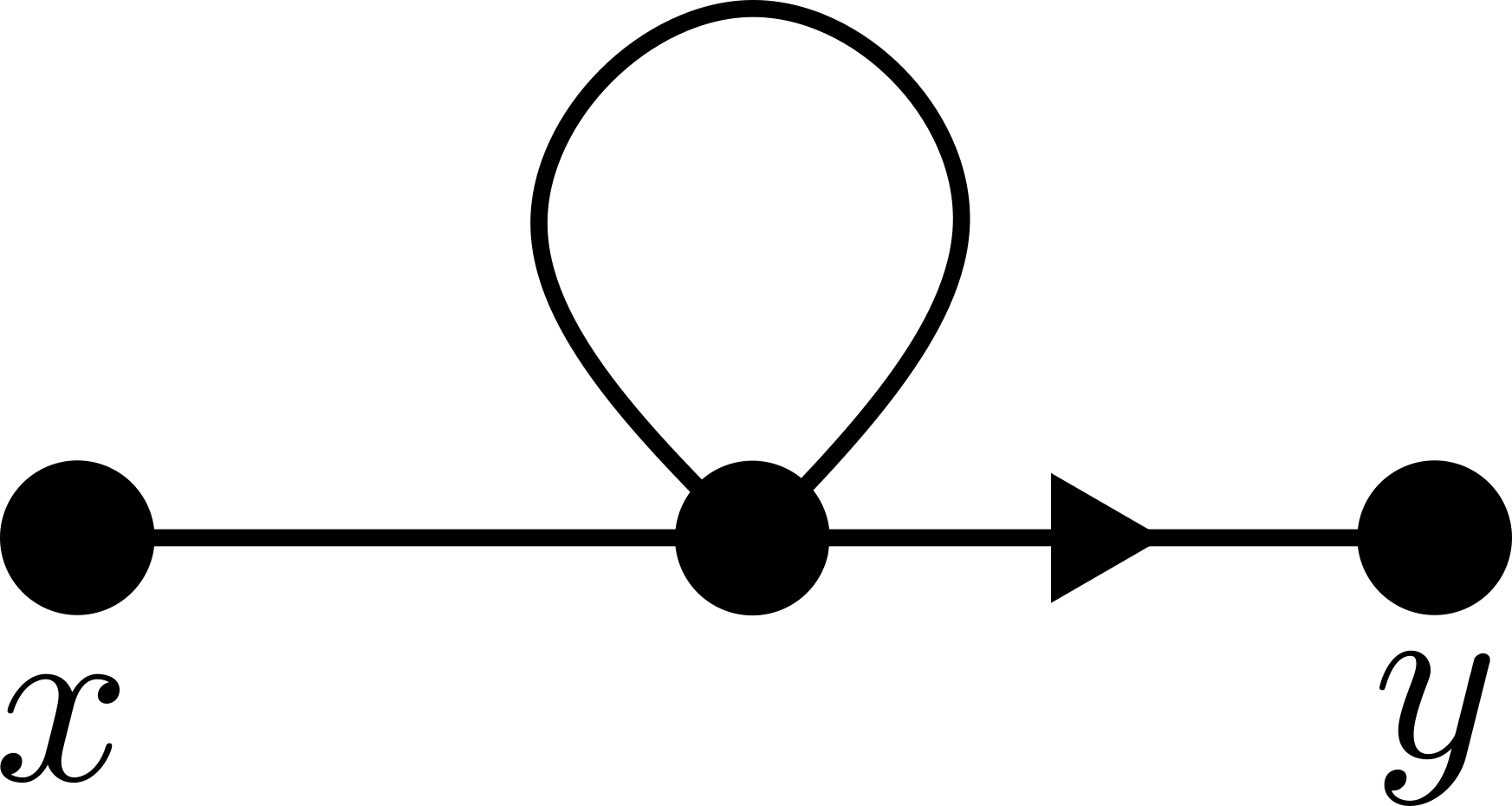} + \diagram{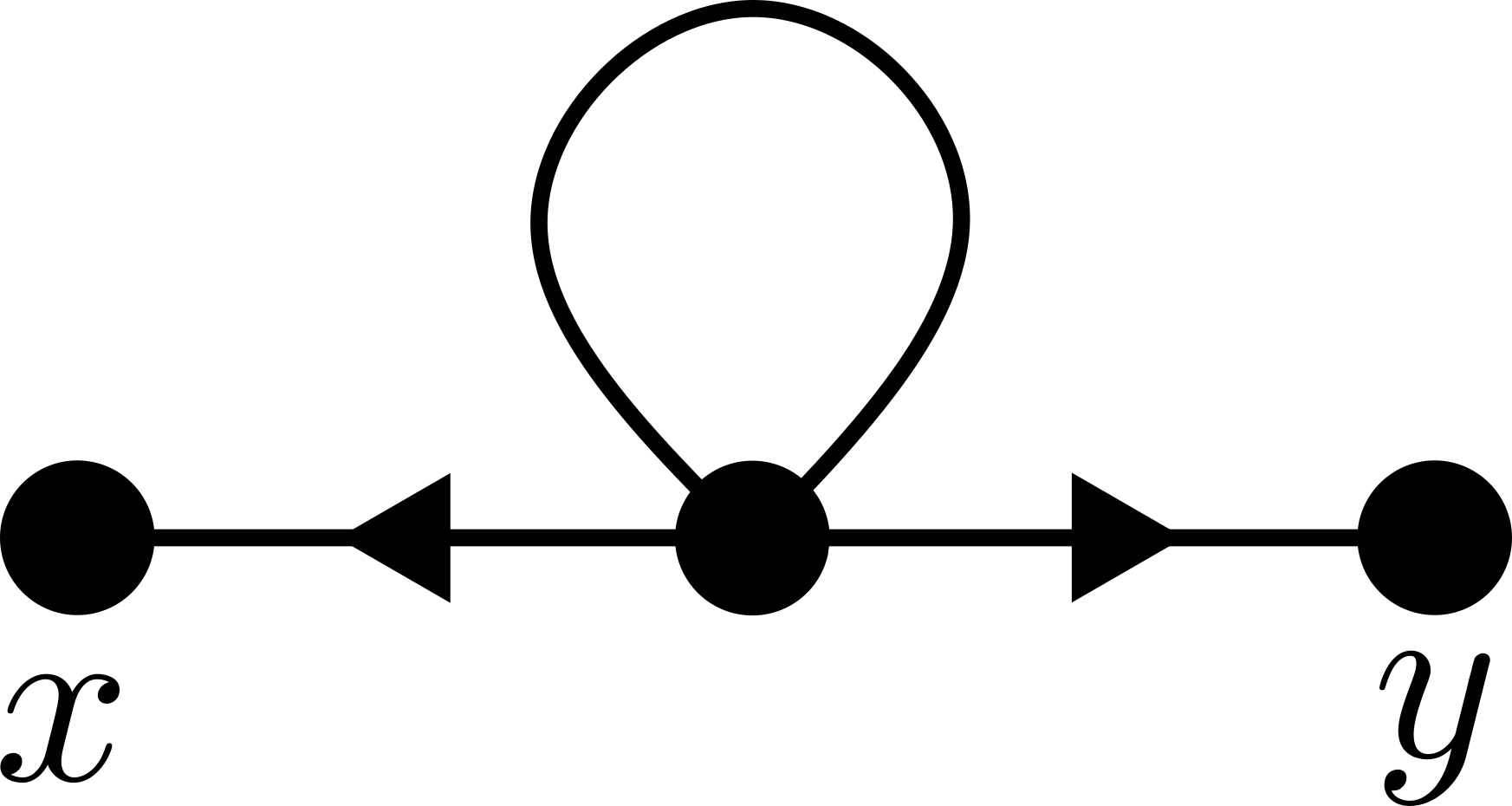} + \; ... \, ,
\end{align}
where the possible diagrams on the rhs consist of all the \textit{connected} graphs one can draw by adding some number of internal vertices with 4 legs each (here we have only shown those with 1 internal vertex for brevity). One then `dresses' the graphs with directed edges in all possible ways such that i) by following directed edges every internal vertex has a route out to either $x$ or $y$, or both, and ii) every such route leads to $x$ or $y$ (thus, there are no closed loops of directed edges).

The contribution of a given diagram can be computed from the following rules:
\begin{enumerate}
    \item For each internal vertex $z$ we get a sum $\frac{i}{\rho}\sum_{z \in C} \lambda_{z}$. \hspace{32.5mm}\diagram{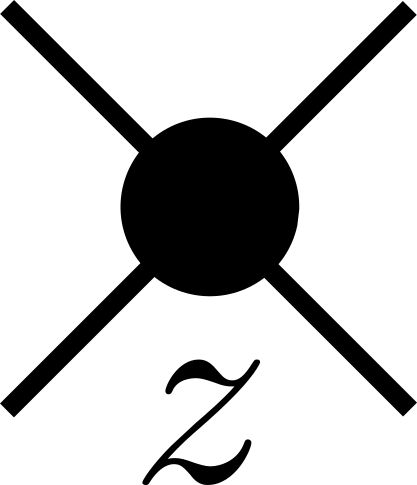}

    \item For each directed leg from a vertex $a$ to $b$ we get a factor $i G_{ba}$. \hspace{3mm} \diagram{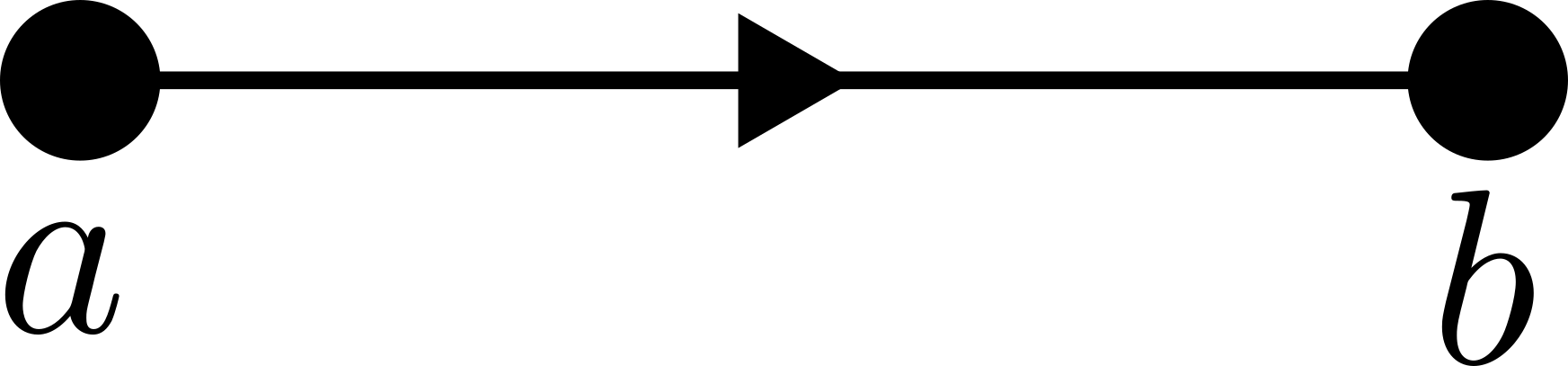}

    \item For each undirected leg from $a$ to $b$ we get a factor $G^F_{ab}$. \hspace{16mm} \diagram{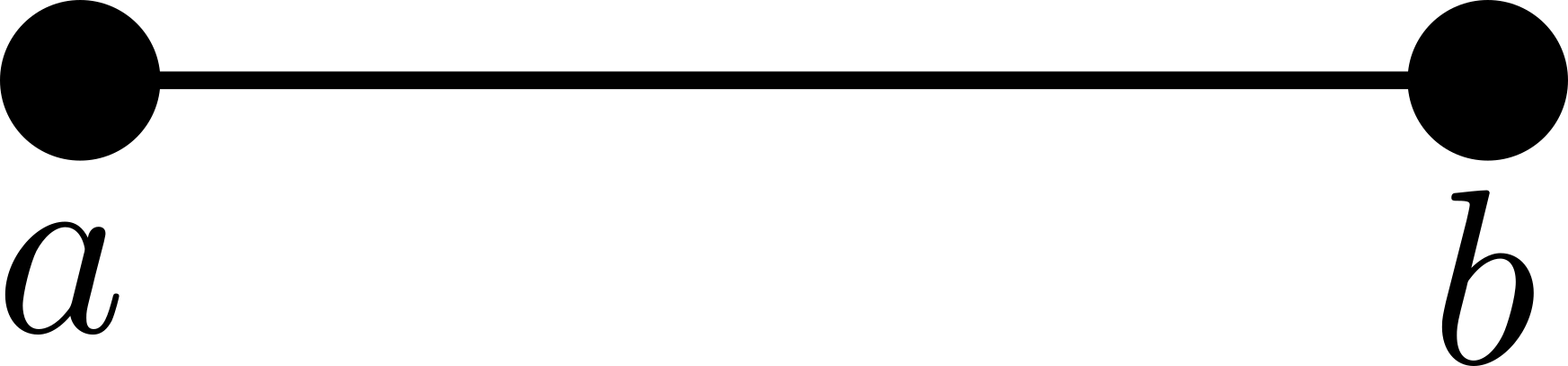}

    \item For each internal vertex $z$, if it has $p$ outgoing directed legs we get a factor of $(-1)^{p-1}$.
    
    \item Divide by symmetry factor, $\mathtt{S}_D$, of the diagram $D$. Equivalently, multiply by $1/\mathtt{S}_D$.
\end{enumerate}

The symmetry factor, $\mathtt{S}_D$, of a diagram $D$ with $n$ internal vertices, is simply the factor that is left over when we divide $(4!)^n$ by all the numerical factors we get in the construction of $D$ following the steps of the previous section. It represents certain symmetries of the given diagram. For example, $\mathtt{S}_D$ contains a factor of 2 for any loops from a vertex to itself (e.g. the loop from $z_2$ to itself in Figure~\ref{fig:labelleddiagramz1z2example}), as the diagram is symmetric under the interchange of the ends of such a line. $\mathtt{S}_D$ contains a factor of $q!$ if the diagram is symmetric under the interchange of $q$ lines (e.g. the two directed lines from $z_1$ to $z_2$, or the two undirected lines from $z_1$ to $z_3$ in Figure~\ref{fig:finalstep}). Factors also arise if the diagram is symmetric under the interchange of certain vertices. In general, it can be non-trivial to determine the symmetry factor, so we will not go into this here. 

We now describe some notable features of our diagrams. Due to the construction process, all the diagrams are connected. In the continuum, one finds both connected and disconnected diagrams in the numerator and denominator of the rhs of~\eqref{eq:peskin path integral}, and further that only the connected diagrams remain after taking the quotient. In our case, no disconnected diagrams ever arise.

Another interesting feature is that causality is explicitly encoded through the appearance of the retarded Green function. These factors ensure that the correlation between the fields at $x$ and $y$ only differs from the free correlation, $W_{xy}=\bra{\Omega}\phi_x \phi_y\ket{\Omega}$, if $x$ and $y$ sit to the future of some points $z$ for which the interaction is turned on, i.e. $\lambda_z\neq 0$. This manifest causality in our diagrams is reminiscent of~\cite{Dickinson_2014}, though in that case they were concerned with sources and detectors. It would be interesting to see if further comparisons between our diagrams and those of~\cite{Dickinson_2014} can be drawn.

Finally, the above procedure can be immediately generalised from the SJ state to \textit{any} Gaussian state for which $n$-point functions split into sums of products of 2-point functions. This latter feature is all that was really needed from the state in our prescription.

\section{Discussion}

We developed interacting real scalar QFT on a fixed causal set. We did this first via the double path integral framework, following Sorkin's suggestion in~\cite{Sorkin_scalar_field_histories} to modify the analogue of the causal set action to include a self-interacting $\phi^4$ term. We then used our interacting decoherence functional to derive the corresponding modification to the canonical theory. This amounted to a unitary transformation of the field operators at each causal set point, i.e. equation~\eqref{eq:unitary relationship between interaction and heisenberg picture fields}. We further highlighted the similarities between the double path integral and canonical descriptions of our causal set interacting QFT, and those of the continuum. 

After answering some initial questions surrounding the interacting decoherence functional, we focused on the interacting 2-point function. We determined how to compute this 2-point function, order by order in the interaction parameter $\lambda$, using a diagrammatic approach. The diagrams that arose in our case resembled those of the continuum but with both directed and undirected edges. This construction can also be generalised without too many complications to general $n$-point functions.

With this basic framework of interacting QFT on causal sets laid out, there are many interesting directions for future investigations. One could compare the interacting 2-point function of the causal set with that of the continuum. For further comparison, one could use the above framework to determine scattering amplitudes between eigenstates of the free theory, and compare them to the textbook continuum expressions.

One important aspect of continuum interacting QFT, and one not considered here, is that of \textit{renormalisation}. To study this in the framework outlaid above, one would first need to specify some coarse-graining procedure, either by i) removing some causal set points, or by ii) truncating the `higher energy' modes from the eigenspectrum, or by some other means altogether. With some coarse-graining procedure in place, one could then integrate out the appropriate degrees of freedom from the double path integral. For i), one would simply integrate out the values of $\xi_x$ and $\overline{\xi}_x$ for every point $x\in C$ that was thrown away. For ii), one would first expand $\xi$ and $\overline{\xi}$ in the eigenmodes of $i\Delta$, and then integrate out those modes above the cut-off. To study renormalisation one could then look at whether there is some dependence of the coupling $\lambda$ on the cut-off that leaves the interacting $n$-point functions constant wrt the cut-off. Such a dependence could then be interpreted as the running of the coupling $\lambda$.

\bibliography{references}

\end{document}